\documentclass[11pt, oneside]{article}   	

 \usepackage[margin=1.5cm]{geometry}                		
\usepackage{graphicx}				
\usepackage{amssymb}
\usepackage{amsmath}

\title{Force networks, torque balance and Airy stress \\ in the planar vertex model of a confluent epithelium}

\author{Oliver E. Jensen,$^1$\footnote{Oliver.Jensen@manchester.ac.uk}~~
Emma Johns$^2$ \& Sarah Woolner$^2$ \\ \\
$^1$Department of Mathematics, University of Manchester, \\ Oxford Road, Manchester M13 9PL, UK \\
$^2$Wellcome Trust Centre for Cell-Matrix Research, \\ Division of Cell Matrix Biology and Regenerative Medicine, School of Biological Sciences, \\ University of Manchester, Oxford Road, Manchester, M13 9PT, UK}
 \date{\today}							

\begin{document}
 \maketitle


\begin{abstract}
The vertex model is a popular framework for modelling tightly packed biological cells, such as confluent epithelia.   Cells are described by convex polygons tiling the plane and their equilibrium is found by minimizing a global mechanical energy, with vertex locations treated as degrees of freedom.  Drawing on analogies with granular materials, we describe the force network for a localized monolayer and derive the corresponding discrete Airy stress function, expressed for each $N$-sided cell as $N$ scalars defined over kites covering the cell.  We show how a torque balance (commonly overlooked in implementations of the vertex model) requires each internal vertex to lie at the orthocentre of the triangle formed by neighbouring edge centroids.  Torque balance also places a geometric constraint on the stress in the neighbourhood of cellular trijunctions, and requires cell edges to be orthogonal to the links of a dual network that connect neighbouring cell centres and thereby triangulate the monolayer.  
We show how the Airy stress function depends on cell shape when a standard energy functional is adopted, and discuss implications for computational implementations of the model.
\end{abstract}

\section{Introduction}

Multicellular biological tissues have an intrinsically granular structure, associated with the mechanical integrity of individual cells.   While cells may be sufficiently soft for many tissues to deform like continuous media described by smooth strain fields \cite{ambrosi2019},  stress fields can remain heterogeneous \cite{baskin2013} and may display features that are not captured in terms of smoothly varying (homogenized) variables.  Accordingly, the vertex model of tightly-packed cells  \cite{alt2017, brodland2014, farhadifar2007, fletcher2014, fozard2016, guirao2015, ishihara2017, nagai2001, staple2010, weliky1990} has become a popular framework with which to model plant and animal development, particularly of tightly-packed epithelial monolayers.  The vertex model captures cell geometries efficiently, enables straightforward computation that resolves individual cells, and is based on simple mechanical assumptions.  Integrating over regions, it can be used to derive tissue-scale properties such as elastic moduli \cite{ merkel2019, murisic2015, ANB2018b}.  In addition to capturing a jamming/unjamming phase transition, with resistance to shear vanishing as cells lose cortical tension --- a topic of much current attention \cite{bi2015, bi2016, boromand2018, merkel2019, yang2017} --- the vertex model also exhibits inherently discrete mechanical structures (such as force chains and correlated patterns of stress \cite{gao2016, ANB2018a}), which have the potential to influence biological behaviour.   Despite its popularity, however, the mechanical constraints underpinning the vertex model have not yet been fully articulated.

In classical elasticity, materials are defined with respect to a reference state, using a strain energy function defined in terms of strain invariants.  The vertex model differs in using cell area and perimeter as intrinsic measures of shape (for systems such as epithelia that are well described by two-dimensional models), and the concept of a reference state is not employed.  In many ways the manner in which cells pack together under an external load instead resembles a granular material, which can accommodate multiple configurations under given boundary conditions \cite{bi2015a}.   Here we exploit this analogy to identify the force network associated with a planar cell configuration, and derive the corresponding force potential and Airy stress function.  We show that the Airy stress function is defined over kites that tile individual polygonal cells.  Whereas stress components can be expressed as second derivatives of the Airy stress function in a planar elastic material, here stress is constructed using discrete derivatives, as deployed for granular media \cite{degiuli2011, degiuli2014, satake1993} and in models for self-equilibrated frameworks \cite{fraternali2014}.   Accordingly, we exploit some machinery from graph theory and discrete calculus, making extensive use of incidence matrices, which serve as {analogues of finite-}difference (coboundary) operators (or, when transposed, as boundary operators) \cite{desbrun2008, grady2010, tonti2014, lim2019}, while avoiding the full formalism of exterior calculus.

\begin{figure}
\begin{center}
\includegraphics[height=3.5in]{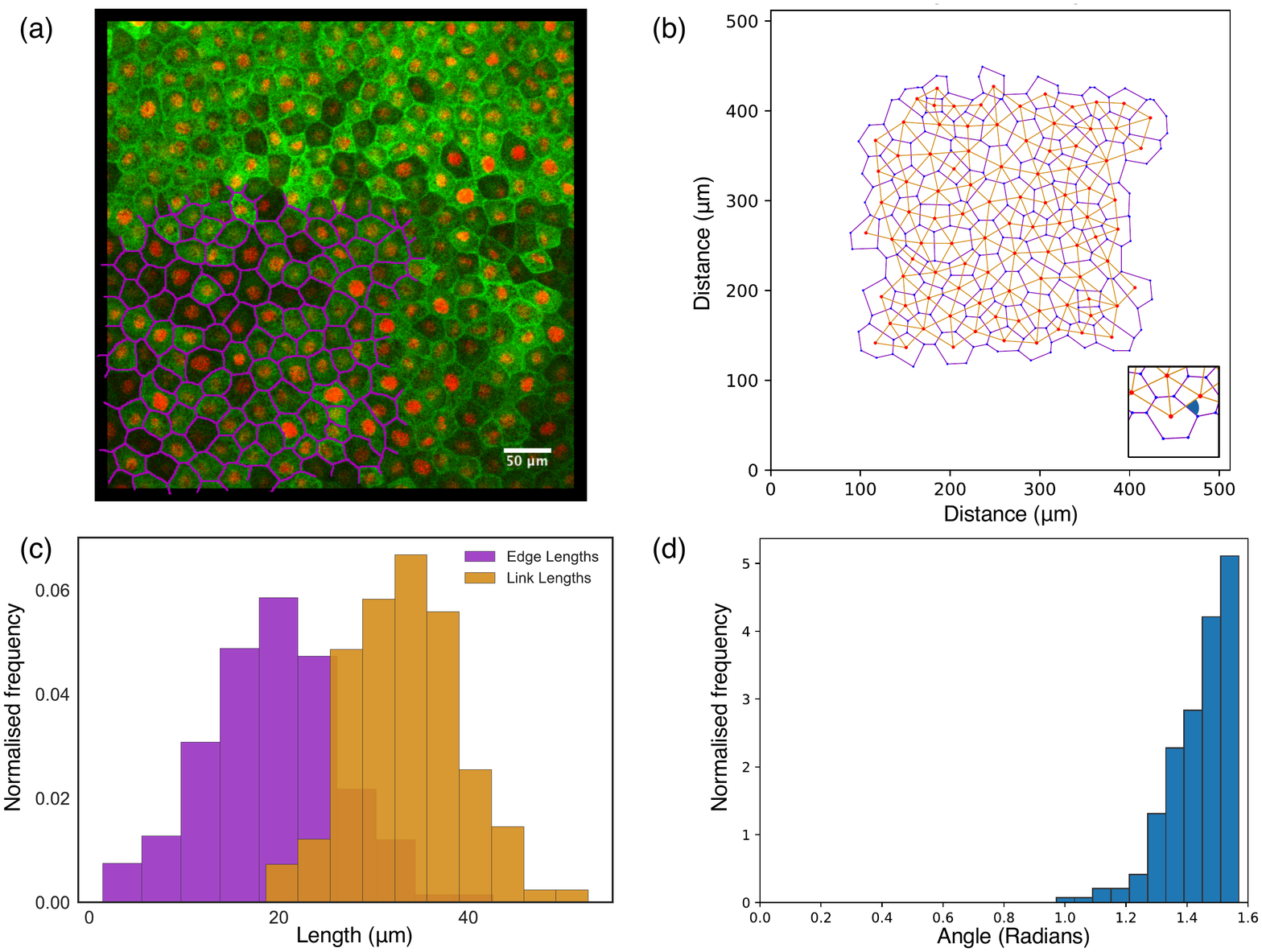}
\end{center}
\caption{(a) An epithelium (animal cap) dissected from a \textit{Xenopus laevis} embryo and adhered to a fibronectin-coated PDMS membrane, imaged by confocal microscopy; cell edges are identified with GFP-alpha-tubulin (green); cell nuclei with cherry-histone 2B (red).   Some cell shapes are mapped out in magenta.  (b) The segmented image, with each cell represented as a polygon bounded by vertices at its trijunctions.  (c) Distributions of edge lengths (between trijunctions) and link lengths (between cell centroids).  (d) The distribution of angles at intersections between links and edges (illustrated by the inset in (b)), peaking at $\pi/2$.}
\label{fig:ex}
\end{figure}

The Airy stress function serves as a discrete scalar potential for the vector force potential, and its existence guarantees that intra- and intercellular stress tensors are symmetric, \hbox{i.e.} that there is a torque (or moment) balance across a monolayer.  We show in the present case that this condition places a geometric constraint on the intercellular stress in the neighbourhood of cellular trijuctions.  This stress-geometry condition is provided by a fabric tensor resembling that described by Ball \& Blumenfeld \cite{ball2002} for granular materials; to our knowledge it has not been used previously in the context of the vertex model.  We show how the fabric tensor can be used to determine the orientation of stress in the neighbourhood of trijuctions.  Furthermore, we show that a torque balance in intercellular stress (not normally considered in biological studies that focus on intracellular stress, nor imposed in simulations that only apply a point-wise force balance on vertices) reveals the requirement that links between cell centres (appropriately defined) should, within the framework of the vertex model, be orthogonal to the cell edges that they intersect and, {crucially,} that each cell vertex should lie at the orthocentre of the triangle connecting adjacent edge centroids.  {We show how these constraints can be used to identify a consistent triangulation of the monolayer that is dual to the network of cell boundaries.}   

The vertex model is of course a simple idealisation of a complex biological system.   The geometry of a typical epithelium (Fig.~\ref{fig:ex}a) is summarised by the locations of its trijunctions (vertices), combined with topological information identifying the cell edges connecting vertices, and the cells that are bounded by edges (Fig.~\ref{fig:ex}b).  This {primal} cellular network generically shows a degree of intrinsic disorder, captured for example by a distribution of edge lengths (Figure~\ref{fig:ex}c).  Figure~\ref{fig:ex}(b) illustrates one possible {dual} network, constructed in this instance by links connecting the centroids (defined with respect to cell vertices) of adjacent cells.  The links also show variability in length (Figure~\ref{fig:ex}b).  The angles at which links intersect their corresponding cell edges are quite tightly distributed around $\pi/2$ (Figure~\ref{fig:ex}d), but show some evidence of non-orthogonality.  We discuss this observation in light of theoretical predictions below.

In this study we ignore neighbour exchanges (T1 transitions), cell extrusion, cell division and intrinsic cell motility, focusing simply on monolayer configurations with fixed topology.  {For simplicity, we also assume that all internal vertices are trijunctions.}  In Sec.~\ref{sec:dis} we implement the planar vertex model using incidence matrices and lay out some relevant geometric and topological results before representating intra- and intercellular stress fields in terms of potentials in Sec.~\ref{sec:str}.   These results are intrinsic to the vertex-based description and independent of a constitutive model, which we introduce in Sec.~\ref{sec:con}.   Adopting a widely used approximation for cell elastic energy, we show how intracellular variations in Airy stress function are proportional to the cell's cortical tension, and can be expressed directly in terms of cell shape.  Findings are summarised in Sec.~\ref{sec:disc}, where we {propose} a potential computational strategy that respects torque balance and { discuss} the relevance of the model to real epithelia.

\section{The planar vertex model}
 \label{sec:dis}

We consider a localized monolayer of $N_c$ confluent cells, represented as tightly-packed polygons covering a simply connected region of the plane.  We assume that an external isotropic stress $P_{\mathrm{ext}}$ is applied around the periphery of the monolayer.   In computations, starting from some (typically disordered) initial condition, vertex locations either evolve under a local force balance until the system reaches equilibrium, or they are adjusted directly to minimize a global energy.   In either case, each vertex in the monolayer can be assumed instantaneously to be under zero net force (inertial effects are neglected).  We wish to understand the impact of imposing, additionally, a torque balance across the monolayer.

\subsection{Cell topology and geometry}

{Given the nature of the vertex model, and the quality of available imaging data, we take cell boundaries as the primal network, which we assume is embedded in a Euclidean space.}  The cellular monolayer is {therefore} defined by a set of vertices (position vectors) $\mathbf{r}_k\in \mathbb{R}^2$, $k=1,\dots, N_v$, a set of oriented cell edges $\mathbf{t}_j$ (of length $t_j$), $j=1,\dots, N_e$ and a set of oriented cell faces (that we simply call \textit{cells}) $\mathsf{a}_i$   (of area $A_i$), $1=1,\dots, N_c$.  Here $\mathsf{a}_i=A_i \boldsymbol{\epsilon}_i$ where $\boldsymbol{\epsilon}_i=\pm \boldsymbol{\varepsilon}$ represents a clockwise rotation by $\pm \pi/2$. ($\boldsymbol{\varepsilon}$ is the 2D Levi-Civita symbol satisfying  $\boldsymbol{\varepsilon}^T=-\boldsymbol{\varepsilon}$,  $\boldsymbol{\varepsilon} \boldsymbol{\varepsilon}=-\mathsf{I}$; the summation convention is not adopted here.).  
Orientations of edges and faces are prescribed but arbitrary; here we will assume that all cells have the same orientation.  We collect vertices, edges and faces into vectors $\mathbf{r}=(\mathbf{r}_1,\dots,\mathbf{r}_{N_v})^T$, $\mathbf{t}=(\mathbf{t}_1,\dots,\mathbf{t}_{N_e})^T$ and $\mathsf{a}=(\mathsf{a}_1,\dots,\mathsf{a}_{N_c})^T$ but for clarity use matrix notation sparingly below, writing sums explicitly in many cases.

The topology of the monolayer is defined using two \textit{incidence matrices} \cite{grady2010}.  The $N_e\times N_v$ matrix $\mathsf{A}$ has elements $A_{jk}$ that equal $1$ (or $-1$) when edge $j$ is oriented into (or out of) vertex $k$, and zero otherwise.  The $N_c\times N_e$ matrix $\mathsf{B}$ has elements $B_{ij}$ that are non-zero only when edge $j$ is on the boundary of cell $i$, taking values $+1$ if the edge is coherent with the orientation of the cell face and $-1$ if not.  Replacing $-1$ with 1 in each matrix produces {unsigned incidence} matrices $\overline{\mathsf{A}}$ and $\overline{\mathsf{B}}$, identifying neighbours but not  orientations.  Further properties of $\mathsf{A}$ and $\mathsf{B}$ are given in Appendix~\ref{sec:bo}.   The $N_c\times N_v$ matrix ${\mathsf{C}}=\tfrac{1}{2} \overline{\mathsf{B}} \, \overline{\mathsf{A}}$ has elements ${C}_{ik}$ that equal $1$ if vertex $k$ neighbours cell $i$ and zero otherwise.   Thus $Z_i\equiv \sum_k {C}_{ik}$ (summing over all vertices) defines the number of edges (and vertices) of cell $i$.   We let $\mathbf{R}_i$ represent the centre of each cell, without specifying yet how it might be related to the cell's vertex locations $\cup_k C_{ik}\mathbf{r}_k$ (where $\cup_k$ denotes collection, without summation, over all vertices).
 
\begin{figure}
\begin{center}
\includegraphics[height=3.5in]{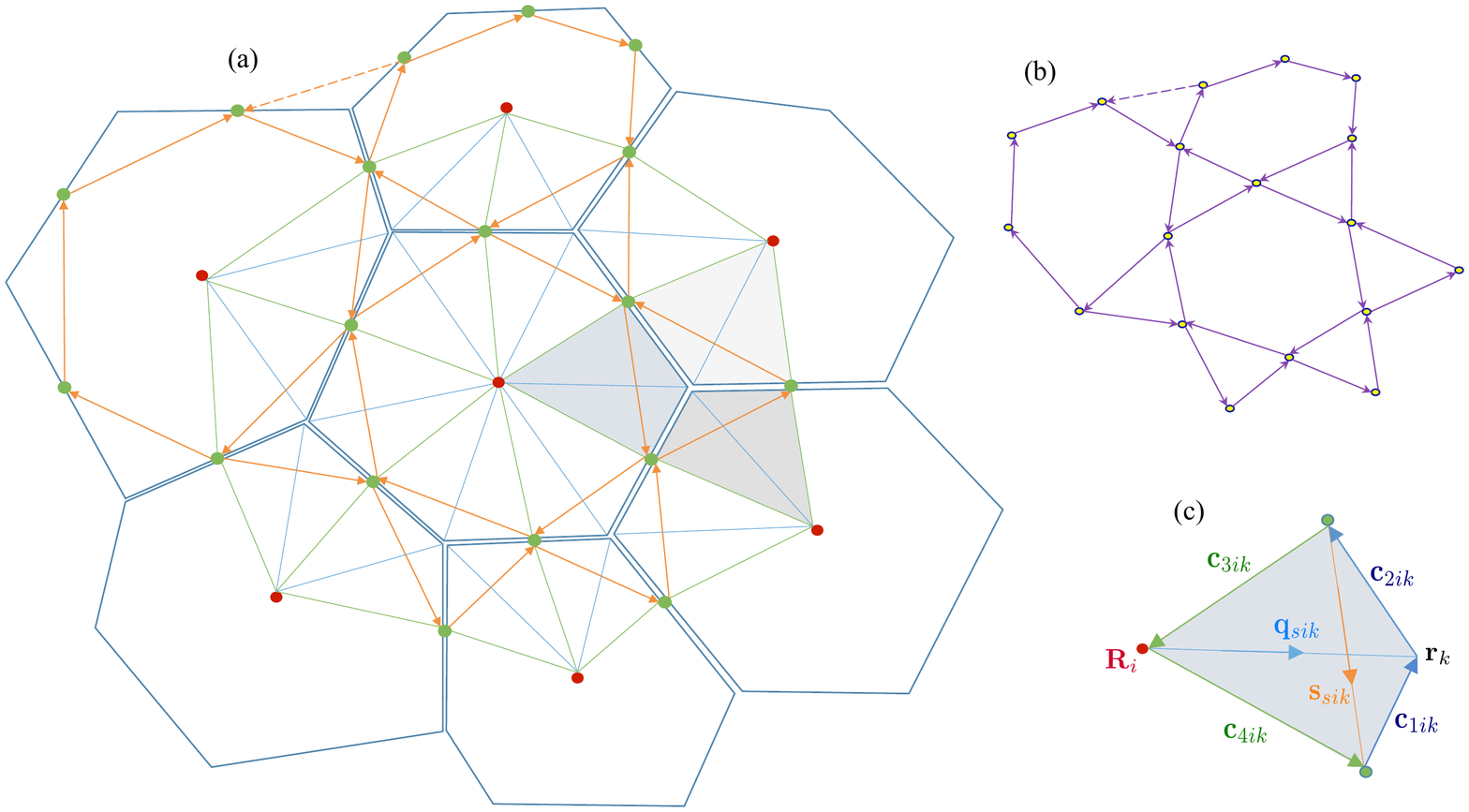}
\end{center}
\caption{(a) An illustration of a localized monolayer. Blue lines show cell edges, meeting at vertices.  This example has $N_c=7$ cells (6 border, 1 interior), $N_e=30$ edges (18 peripheral, 6 border, 6 interior), $N_v=24$ vertices (18 peripheral, 6 interior).  Orientations of edges and faces are not indicated.  Green dots are centroids $\mathbf{c}_j$ of each edge and red dots illustrate centres $\mathbf{R}_i$ of each cell.  The solid orange lines connecting edge centroids form triangles around each internal vertex and polygons around each cell.  Each cell is constructed from \textit{kites}: three kites (shaded) surrounding an internal vertex together define a \textit{tristar}. A force $\mathbf{f}_{ik}$ due to cell $i$ on vertex $k$ is associated with each kite.  (b) Solid purple arrows show rotated forces $-\boldsymbol{\varepsilon}\mathbf{f}_{ik}$.  The force balances on vertices and cells imply that the rotated force vectors form a network that has the topology of the network containing edge centroids.  The centroids $\mathbf{c}_j$ therefore map to vertices of the force network $\mathbf{h}_j$ (circular symbols).  An imposed uniform pressure is represented by the peripheral forces, represented in part by supplementary links (dashed) that close triangles.  (c) Kite $ik$, spanned by the vector $\mathbf{q}_{ik}$ from the centre of cell $i$ to vertex $k$ and the vector $\mathbf{s}_{ik}$ connecting the centroids of the edges adjacent to vertex $k$.  The vectors $\mathbf{c}_{1ik},\dots,\mathbf{c}_{4ik}$ bounding the kite are also indicated. }
\label{fig:regionsA}
\end{figure}

To account for boundaries of the monolayer, vertices (and all other functions defined on vertices, with subscript $k$) are partitioned as $N_p$ peripheral and $N_v-N_p$ interior vertices so that $\mathbf{r}=[\mathbf{r}^p, \mathbf{r}^i]^T$, edges (and relevant functions with subscript $j$) as $N_p$ peripheral, $N_b$ border and $N_e-N_p-N_b$ interior edges so that $\mathbf{t}=[\mathbf{t}^p, \mathbf{t}^b, \mathbf{t}^i]^T$, and cells (and functions with subscript $i$) as $N_b$ border and $N_c-N_b$ interior cells so that $\mathsf{a}=[\mathsf{a}^b, \mathsf{a}^i]^T$.  A peripheral edge has two peripheral vertices; a border edge has one peripheral and one interior vertex; an interior cell has only interior edges.   Internal vertices always represent trijunctions.  Fig.~\ref{fig:regionsA}(a) illustrates this for a small monolayer of 7 cells.  We may then partition the incidence matrices as
\begin{equation}
\mathsf{A}=\left(
\begin{matrix}
\mathsf{A}^{pp} & \mathsf{0} \\
\mathsf{A}^{bp} & \mathsf{A}^{bi} \\
\mathsf{0} & \mathsf{A}^{ii} 
\end{matrix}
\right),\quad
\mathsf{B}=\left(
\begin{matrix}
\mathsf{B}^{bp} & \mathsf{B}^{bb} & \mathsf{B}^{bi} \\
 \mathsf{0} & \mathsf{0} & \mathsf{B}^{ii} 
\end{matrix}
\right)
\end{equation}
where $\mathsf{A}^{pp}$ is an $N_p\times N_p$ matrix, etc., so that
\begin{equation}
\mathsf{C}\equiv  \left(
\begin{matrix}
\mathsf{C}^{bp} & \mathsf{C}^{bi} \\
\mathsf{0} & \mathsf{C}^{ii} 
\end{matrix}
\right)=
\frac{1}{2} \left(
\begin{matrix}
\overline{\mathsf{B}}^{bp} \overline{\mathsf{A}}^{pp} + \overline{\mathsf{B}}^{bb} \overline{\mathsf{A}}^{bp} &  
\overline{\mathsf{B}}^{bb} \overline{\mathsf{A}}^{bi} + \overline{\mathsf{B}}^{bi} \overline{\mathsf{A}}^{ii} \\
\mathsf{0} & \overline{\mathsf{B}}^{ii} \overline{\mathsf{A}}^{ii}
\end{matrix}
\right).
\label{eq:Cik}
\end{equation}

Edges are defined by $\mathbf{t}_j=\sum_k A_{jk}\mathbf{r}_k$, with lengths $t_j=\sqrt{ \mathbf{t}_j \cdot \mathbf{t}_j}$.   
This defines the unit vectors $\hat{\mathbf{t}}_j=\mathbf{t}_j/t_j$.  
The perimeter of cell $i$ is $L_i=\sum_j \overline{B}_{ij}t_j$ (summing over all edges).  It follows (for later reference) that
\begin{equation}
\frac{\partial t_j}{\partial \mathbf{r}_k}=\hat{\mathbf{t}}_jA_{jk} \quad\mathrm{and}\quad
\frac{\partial L_i}{\partial \mathbf{r}_k}={\textstyle{\sum_j}} \overline{B}_{ij} \hat{\mathbf{t}}_j A_{jk}.
\label{eq:perim}
\end{equation}
$\partial L_i/\partial \mathbf{r}_k$ is therefore the sum of two unit vectors aligned with the two edges of cell $i$ that meet vertex $k$, pointing into the vertex. 

To define cell areas, we construct 
\begin{equation}
\mathbf{n}_{ij}=-\boldsymbol{\epsilon}_i B_{ij} \mathbf{t}_j \quad\mathrm{and}\quad \mathbf{c}_j=\tfrac{1}{2}{\textstyle{\sum_k}} \overline{A}_{jk}\mathbf{r}_k.
\end{equation}
$\mathbf{n}_{ij}$ defines the outward normal of cell $i$ at edge $j$ and $\mathbf{c}_j$ defines the centroid of edge $j$.  Let $\phi=\tfrac{1}{2}\mathbf{x}\cdot\mathbf{x}$ where $\mathbf{x}$ is a position vector in $\mathbb{R}^2$ and integrate $(\nabla\otimes \nabla) \phi=\nabla\otimes \mathbf{x}=\mathsf{I}$ over cell $i$, where $\otimes$ denotes the dyadic outer product.  Applying the divergence theorem to an integral over cell $i$,
\begin{equation}
A_i\mathsf{I}=\int_i  \nabla\otimes\nabla \phi \,\mathrm{d}A= \oint_{\partial i} \hat{\mathbf{n}}\otimes \mathbf{x} \, \mathrm{d}s={\textstyle{\sum_j}} \mathbf{n}_{ij} \otimes \mathbf{c}_j \equiv {\textstyle{\sum_{j} }} \mathbf{n}_{ij} \mathbf{c}_j^T.
\label{eq:aten}
\end{equation}
The oriented area of cell $i$ can therefore be written as 
\begin{equation}
\mathsf{a}_i\equiv A_i \boldsymbol{\epsilon}_i={\textstyle{\sum_j}} B_{ij} \mathbf{t}_j\otimes \mathbf{c}_{j}.
\label{eq:a2}
\end{equation}
The trace of (\ref{eq:a2}) gives $\sum_j B_{ij}\mathbf{t}_j\cdot \mathbf{c}_j=0$. 
This can be understood by recognising $\phi$ as the potential for position $\mathbf{x}=\nabla \phi$; its discrete form $\phi_k=\phi(\mathbf{r}_k)$ jumps by $\sum_k A_{jk}\phi_k=\mathbf{c}_j\cdot \mathbf{t}_j$ along edge $j$, and the net change in potential vanishes around a closed loop because $\mathsf{B}\mathsf{A}=\mathsf{0}$ (Appendix~\ref{sec:bo}), a device we will exploit later on.  Also, as shown elsewhere (\hbox{e.g.} \cite{ANB2018a, yang2017}),
\begin{equation}
\frac{\partial A_i}{\partial \mathbf{r}_k}
=\tfrac{1}{2} {\textstyle{\sum_j}} \boldsymbol{\epsilon}_i B_{ij}   \mathbf{t}_j\overline{A}_{jk} \equiv -\tfrac{1}{2}{\textstyle{\sum_j}} \mathbf{n}_{ij} \overline{A}_{jk}.
\label{eq:area}
\end{equation}
$\partial A_i/\partial\mathbf{r}_k$ is therefore the sum of two inward normal vectors associated with the edges of cell $i$ meeting at vertex $k$, with length equal to half of each edge.  

\subsection{Dual networks}

There are multiple networks that are dual to the (primal) network of cells.  The simplest is the triangulation (a simplicial complex) connecting adjacent cell centres.  Assigning orientation $\boldsymbol{\epsilon}_{k}$ to all triangles (opposite to that in all cells), the orientations of links between cell centres are induced by the choice of $\mathsf{A}$ and $\mathsf{B}$ (Appendix~\ref{sec:bo}), with link $\mathbf{T}_j=\sum_i B_{ij}\mathbf{R}_i$ dual to edge $\mathbf{t}_j$.  For a localized monolayer, peripheral triangles and links are truncated; complete links are given by $\mathbf{T}^b=\mathsf{B}^{bbT} \mathbf{R}^b$ and $\mathbf{T}^i=\mathsf{B}^{biT}\mathbf{R}^b+\mathsf{B}^{iiT}\mathbf{R}^i$, where $\mathbf{T}=(\mathbf{T}_1,\dots,\mathbf{T}_{N_e})^T=[\mathbf{T}^p,\mathbf{T}^b,\mathbf{T}^i]^T$ and $\mathbf{R}=(\mathbf{R}_1,\dots,\mathbf{R}_{N_v})^T=[\mathbf{R}^p,\mathbf{R}^i]^T$.

We will also make use of a second dual network, formed by links between cell centres and edge centroids (Fig.~\ref{fig:regionsA}a).  This partitions each cell into \textit{kites} (described in more detail below), with three kites surrounding each vertex.  The resulting 6-sided \textit{tristar} at each vertex shares 3 vertices with the triangle connecting cell centoids, but their edges in general are distinct.  We denote the area of the tristar at vertex $k$ as $E_k$.


A more fine-grained \textit{edge-centroid} network is built by connecting adjacent edge centroids around each cell.  Thus 
\begin{equation}
\mathbf{s}_{ik}=\boldsymbol{\varepsilon}{\textstyle{\sum_j}}  \boldsymbol{\epsilon}_i B_{ij} \mathbf{c}_j A_{jk}
\label{eq:cik}
\end{equation}
defines links between adjacent edge centroids running clockwise as polygons around cells and anticlockwise as triangles around vertices (Fig.~\ref{fig:regionsA}a; Appendix~\ref{sec:cik}).  To invert (\ref{eq:cik}) we may use
\begin{equation}
\mathbf{c}_j-\mathbf{c}_{j'}=\sum_{ik\in \mathcal{P}(j,j')} \mathbf{s}_{ik}
\label{eq:cj}
\end{equation}
where $\mathcal{P}(j,j')$ denotes the set of paths over the edge-centroid network connecting $j$ and $j'$, demonstrating how $\mathbf{c}_j$ is a discrete vector potential for $\mathbf{s}_{ik}$.  As loops around any interior vertex $k$ or any cell $i$ are closed, it follows that
\begin{equation}
{\textstyle{\sum_i}} {C}_{ik}\mathbf{s}_{ik}=\mathbf{0}, \quad {\textstyle{\sum_k}} {C}_{ik} \mathbf{s}_{ik}=\mathbf{0}.
\label{eq:edcent}
\end{equation}
More generally, the $N_c\times N_v$ matrix $\mathsf{S}$ with elements $\{\mathsf{S}\}_{ik}=\mathbf{s}_{ik}$ can be combined with $\mathsf{C}$ in (\ref{eq:Cik}) to give $\mathrm{tr}(\mathsf{C}^{bpT}\mathsf{S}^{bp})=0$, because the boundary of the centroid network is closed, while diagonal elements of $\mathsf{C}^{biT}\mathsf{S}^{bi}+\mathsf{C}^{iiT}\mathsf{S}^{ii}$ vanish (representing closed loops around interior vertices); all diagonal elements of $\mathsf{C}\mathsf{S}^T$ vanish (representing closed loops around cells).

Finally, dual to the edge-centroid network is the network of spokes connecting cell centres to vertices.  The outward radial spokes of cell $i$ satisfy $C_{ik}(\mathbf{q}_{ik}-\mathbf{r}_k+\mathbf{R}_i)=\mathbf{0}$ (Fig.~\ref{fig:regionsA}c).

\subsection{Kites}

We combine spokes and links between edge-centroids to build {kites} (Fig.~\ref{fig:regionsA}a,c).  
The links between the cell centre and the edge centroids defining the boundaries of kite $ik$ within cell $i$ are
\begin{equation}
\mathbf{c}_{3ik}=\mathbf{R}_i-\tfrac{1}{2}(\mathbf{u}_{ik}-\mathbf{s}_{ik}),\quad 
\mathbf{c}_{4ik}=\tfrac{1}{2}(\mathbf{u}_{ik}+\mathbf{s}_{ik})-\mathbf{R}_i,
\label{eq:c3c4}
\end{equation}
where $\mathbf{u}_{ik}=\sum_j \overline{B}_{ik} \mathbf{c}_j \overline{A}_{jk}$ is the sum of the two edge centroids bounding kite $ik$, so that $\mathbf{c}_{3ik}$ and $\mathbf{c}_{4ik}$ run anticlockwise around the kite (Fig.~\ref{fig:regionsA}c).  The area of the kite is given by $K_{ik}\boldsymbol{\varepsilon}=\tfrac{1}{2}(\mathbf{s}_{ik} \otimes \mathbf{q}_{ik} - \mathbf{q}_{ik}\otimes \mathbf{s}_{ik})$ (see Appendix \ref{sec:kites}).      Following \cite{ball2002}, we can write
\begin{equation}
\mathbf{s}_{ik}\otimes \mathbf{q}_{ik} =  K_{ik}\boldsymbol{\varepsilon} - \tfrac{1}{2} (\mathbf{c}_{1ik} \otimes \mathbf{c}_{1ik} -\mathbf{c}_{2ik} \otimes \mathbf{c}_{2ik}+\mathbf{c}_{3ik} \otimes \mathbf{c}_{3ik}-\mathbf{c}_{4ik} \otimes \mathbf{c}_{4ik}),
\end{equation}
where $\mathbf{c}_{1ik}$ and $\mathbf{c}_{2ik}$ run anticlockwise along cell edges.
The area of tristar $k$ is therefore $E_k \equiv \sum_i {C}_{ik} K_{ik}$.  Summing kites over the tristar, the internal edge contributions (involving $\mathbf{c}_{1ik}$ and $\mathbf{c}_{2ik}$) cancel leaving only boundary contributions, giving
\begin{equation}
{\textstyle{\sum_i}} {C}_{ik} \mathbf{s}_{ik} \otimes \mathbf{q}_{ik} =  \boldsymbol{\varepsilon} E_k+\mathsf{F}_k, \quad \mathsf{F}_k\equiv  - \tfrac{1}{2} {\textstyle{\sum_i}} {C}_{ik} (\mathbf{c}_{3ik} \otimes \mathbf{c}_{3ik} - \mathbf{c}_{4ik}\otimes \mathbf{c}_{4ik}).
\label{eq:2chaintri}
\end{equation}
The \textit{fabric tensor}  $\mathsf{F}_k$ measures the asymmetry of each tristar \cite{ball2002}; it can be written (Appendix~\ref{sec:cik}) as
\begin{equation}
 \mathsf{F}_k=-\tfrac{1}{2} \boldsymbol{\varepsilon}\boldsymbol{\epsilon}_k\sum_{i,j}  B_{ij}A_{jk}\mathbf{w}_{ij} \otimes \mathbf{w}_{ij}, \quad\mathrm{where}\quad \mathbf{w}_{ij}=\mathbf{c}_j-\mathbf{R}_i.
 \label{eq:fabric}
 \end{equation}
Constructing a cell from kites, edge contributions cancel as well (because kites are defined on edge centroids), giving an alternative formulation of the cell area as
\begin{equation}
\mathsf{a}_i\equiv A_i  \boldsymbol{\varepsilon} = {\textstyle{\sum_k}} {C}_{ik} \mathbf{s}_{ik} \otimes \mathbf{q}_{ik}.
\label{eq:kites}
\end{equation}

\section{Representations of cell and tissue stress}
\label{sec:str}

Let $\mathbf{f}_{ik}$ be the force on vertex $k$ due to cell $i$.  The requirement that the net force at interior vertex $k$ and the net force on any cell $i$ both vanish is
\begin{equation}
{\textstyle{\sum_i} } {C}_{ik}\mathbf{f}_{ik}=\mathbf{0}, \quad {\textstyle{\sum_k}} {C}_{ik}\mathbf{f}_{ik}=\mathbf{0},
\label{eq:foreq}
\end{equation}
representing two {discrete divergences} of $\mathbf{f}_{ik}$.  
 Stating (\ref{eq:foreq}) more generally to account for boundary forcing, we require the diagonal entries of $\mathsf{C}^T (\mathsf{F}-\mathsf{F}_{\mathrm{ext}})$ to vanish (balancing forces at each vertex, including the periphery), and the diagonal entries of $\mathsf{C}\mathsf{F}^T$ to vanish (an internal force balance on each cell), where the matrices $\mathsf{F}$ and $\mathsf{F}_{\mathrm{ext}}$ share the structure of $\mathsf{C}$ in (\ref{eq:Cik}) and $\{\mathsf{F}\}_{ik}\equiv \mathbf{f}_{ik}$.  Now
\begin{equation}
\mathsf{C}^T\mathsf{F}=
\left(\begin{matrix}
\mathsf{C}^{bpT}\mathsf{F}^{bp} & \mathsf{C}^{bpT}\mathsf{F}^{bi} \\
\mathsf{C}^{biT}\mathsf{F}^{bp} & \mathsf{C}^{biT}\mathsf{F}^{bi}+ \mathsf{C}^{iiT}\mathsf{F}^{ii} 
\end{matrix}\right),
\quad
\mathsf{C}\mathsf{F}^T=
\left(\begin{matrix}
\mathsf{C}^{bp}\mathsf{F}^{bpT} + \mathsf{C}^{bi}\mathsf{F}^{biT} & \mathsf{C}^{biT}\mathsf{F}^{iiT} \\
\mathsf{C}^{ii}\mathsf{F}^{biT} & \mathsf{C}^{ii}\mathsf{F}^{iiT}
\end{matrix}\right),
\end{equation}
summing over cells and vertices respectively, and the external force (imposed pressure around the monolayer periphery) has matrix blocks
\begin{equation}
\mathsf{F}_{\mathrm{ext}}^{bp}=\tfrac{1}{2} \mathsf{P}_{\mathrm{ext}}^b \mathsf{B}^{bp}\mathsf{T}^p \overline{\mathsf{A}}^{pp}, \quad
\mathsf{F}_{\mathrm{ext}}^{bi}=\mathsf{0}, \quad
\mathsf{F}_{\mathrm{ext}}^{ii}=\mathsf{0},
\label{eq:fext}
\end{equation}
where $\mathsf{P}^b_{\mathrm{ext}}=P_{\mathrm{ext}}\mathrm{diag}(\boldsymbol{\epsilon}_i^b)$ and $\mathsf{T}^p=\mathrm{diag}(\mathbf{t}_j^p)$.   Thus zero diagonal entries of $\mathsf{C}^{biT}\mathsf{F}^{bi}+\mathsf{C}^{iiT}\mathsf{F}^{ii}$ give (triangular) force balances at interior vertices (including contributions from peripheral cells where appropriate).  Zero diagonal entries of $\mathsf{C}\mathsf{F}^{T}$ give (polygonal) force balances over interior and peripheral cells, and zero diagonals of $\mathsf{C}^{bpT}(\mathsf{F}^{bp}-\mathsf{F}_{\mathrm{ext}}^{bp})$ give the force balance on peripheral vertices.
 

For a monolayer satisfying (\ref{eq:foreq}), the first moment of the force defines the stress $\boldsymbol{\sigma}_i$ associated with cell $i$ via 
\begin{equation}
A_i \boldsymbol{\sigma}_i\equiv {\textstyle{\sum_k}} {C}_{ik} \mathbf{q}_{ik}\otimes \mathbf{f}_{ik} = {\textstyle{\sum_k}} {C}_{ik} \mathbf{r}_k\otimes \mathbf{f}_{ik}.
\label{eq:sigA}
\end{equation}
We call the isotropic component of the stress in each cell the \textit{effective pressure}, $P_{\mathrm{eff},i}\equiv \tfrac{1}{2}\mathrm{tr}(\boldsymbol{\sigma}_i)$.
The stress $\boldsymbol{\sigma}$ of the monolayer as a whole may then be written as
\begin{align}
A\boldsymbol{\sigma}&={\textstyle{\sum_i}} A_i \boldsymbol{\sigma}_i=\sum_{i,k} C_{ik} \mathbf{r}_k\otimes \mathbf{f}_{ik} 
=\sum_{i,k} C_{ik}^{bp} \mathbf{r}^p_k \otimes \mathbf{f}^{bp}_{ik},
\end{align}
where $A=\sum_i A_i$, restricting the final sum to peripheral cells and peripheral vertices because interior forces balance via (\ref{eq:foreq}).  Imposing the boundary condition (\ref{eq:fext}) gives the conservation law \cite{bi2015a}
\begin{equation}
A\boldsymbol{\sigma}=\sum_{i,k} C_{ik}^{bp} \mathbf{r}^p_k \otimes \mathbf{f}^{bp}_{ext, ik} = 
\sum_{i,k} C_{ik}^{bp} \mathbf{r}^p_k \otimes \left(\tfrac{1}{2}P_{\mathrm{ext}} \boldsymbol{\epsilon}_i^b B_{ij}^{bp} \mathbf{t}^p_j  \overline{A}_{jk}^{pp}\right ) 
= - P_{\mathrm{ext}} \sum_{i,k} \mathbf{c}^p_j \otimes \mathbf{n}_{ij}^{bp}   =-A\mathsf{I} P_{\mathrm{ext}},
\label{eq:px}
\end{equation}
showing that the total stress must be isotropic, internal shear stresses must cancel and therefore that \cite{ANB2018a}
\begin{equation}
{\textstyle{\sum_i}} A_i P_{\mathrm{eff},i}=AP_{\mathrm{ext}}.
\label{eq:awp}
\end{equation}
Equation (\ref{eq:px}) also ensures zero net torque on the monolayer due to $P_{\mathrm{ext}}$.

We now consider how the force balances (\ref{eq:foreq}) can be represented geometrically, with a view to identifying the (intercellular) stress $\boldsymbol{\sigma}_k$ defined over tristars.  

\subsection{The force network}

The connection between the force network and the edge centroid network becomes clear if we rotate each force anticlockwise by $\pi/2$ (via a Maxwell--Cremona construction \cite{bi2015a}): then $\sum_i {C}_{ik} (-\boldsymbol{\varepsilon} \mathbf{f}_{ik})=\mathbf{0}$ and  $\sum_k {C}_{ik} (-\boldsymbol{\varepsilon} \mathbf{f}_{ik})=\mathbf{0}$, implying that the rotated force vectors form a network that is topologically equivalent to the edge-centroid network (\ref{eq:edcent}), with closed triangles around vertices and closed polygons around each cell (Fig.~\ref{fig:regionsA}b). 
While the edge-centroid network is planar (by construction), the force network may not be.  In particular, the peripheral forces (\ref{eq:fext}) map to 
\begin{equation}
\tfrac{1}{2} P_{\mathrm{ext}} \boldsymbol{\varepsilon}\mathbf{n}_{ij}^{bp}\overline{A}_{jk}^{pp}
\label{eq:perif}
\end{equation}
which collectively form a closed loop, matching the shape of the perimeter of the edge-centroid network (connecting all the peripheral centroids).  Fixing the location of one peripheral edge centroid at the origin, the loop is clockwise if $P_{\mathrm{ext}}>0$, anticlockwise if $P_{\mathrm{ext}}<0$ and collapses onto the origin if $P_{\mathrm{ext}}=0$.

The centroids $\mathbf{c}_j$ form a discrete potential for the edges $\mathbf{s}_{ij}$ via (\ref{eq:cik}, \ref{eq:cj}).  Similarly, we can identify the vertices of the force network $\mathbf{h}_j$ (Fig.~\ref{fig:regionsA}b) as a potential for the forces,  by writing
\begin{equation}
\mathbf{f}_{ik}=-{\textstyle{\sum_j}} \boldsymbol{\epsilon}_i B_{ij}\mathbf{h}_j A_{jk}, \quad
\mathbf{h}_j-\mathbf{h}_{j'}=\sum_{ik\in\mathbf{P}{(j,j')}} -\boldsymbol{\varepsilon}\mathbf{f}_{ik}.
\label{eq:fint}
\end{equation}
The stress over cell $i$ can then be written in terms of the force potential as
\begin{align}
A_i \boldsymbol{\sigma}_i&={\textstyle{\sum_k}} {C}_{ik} \mathbf{r}_k \otimes \mathbf{f}_{ik} 
=-\sum_{j,k}  \mathbf{r}_k \otimes \left( \boldsymbol{\epsilon}_i B_{ij} \mathbf{h}_j A_{jk} \right) 
 =-{\textstyle{\sum_{j}}} \mathbf{t}_j \otimes \left( \boldsymbol{\epsilon}_i B_{ij} \mathbf{h}_j \right),
\end{align}
noting that $C_{ik}$ becomes redundant when $A_{jk}$ and $B_{ij}$ both appear in the sum, and using $\mathbf{t}_j=\sum_k A_{jk}\mathbf{r}_k$.  Taking a transpose gives 
\begin{equation}
A_i \boldsymbol{\sigma}_i^T =-{\textstyle{\sum_{j}}}  \boldsymbol{\epsilon}_i B_{ij} \mathbf{h}_j \otimes  \mathbf{t}_j.
\label{eq:cellstr}
\end{equation}
$\boldsymbol{\sigma}_i$ should be symmetric for cell $i$ to be under zero torque.  This requires 
\begin{equation}
0 ={\textstyle{\sum_{j}} }  B_{ij} \mathbf{h}_j \cdot  \mathbf{t}_j
\label{eq:cellstr1}
\end{equation}
and allows us to write the cell stress as a discrete curl of $\mathbf{h}$ around its periphery via
\begin{equation}
A_i \boldsymbol{\sigma}_i =-{\textstyle{\sum_{j} } }\boldsymbol{\epsilon}_i B_{ij} \mathbf{h}_j \otimes  \mathbf{t}_j =
{\textstyle{\sum_{j} }} B_{ij} \left( \mathbf{t}_j   \otimes \mathbf{h}_j \right)  \boldsymbol{\epsilon}_i.
\label{eq:cellstr2}
\end{equation}

Likewise $A_i \boldsymbol{\epsilon}_i \boldsymbol{\sigma}_i = \sum_j \mathbf{n}_{ij} \otimes (\boldsymbol{\epsilon}_i \mathbf{h}_j)$, so that $A_i \boldsymbol{\sigma}_i = -\sum_j (\boldsymbol{\epsilon}_i \mathbf{n}_{ij}) \otimes (\boldsymbol{\epsilon}_i \mathbf{h}_j)$, giving the stress in rotated basis.   It follows that
\begin{equation}
A_i \boldsymbol{\epsilon}_i \boldsymbol{\sigma}_i \boldsymbol{\epsilon}_i={ \textstyle{\sum_j} }\mathbf{n}_{ij} \otimes  \mathbf{h}_j.
\label{eq:strrot}
\end{equation}
Comparison with $A_i \mathsf{I}=\sum_j \mathbf{n}_{ij}\otimes \mathbf{c}_j$ (from (\ref{eq:a2})) suggests that stress can be associated with a mapping from $\mathbf{c}_j$ to $\mathbf{h}_j$.  Eq.~(\ref{eq:strrot}) also shows that the isotropic component of cell stress is a discrete divergence, 
\begin{equation}
P_{\mathrm{eff},i}=-\frac{1}{2A_i} {\textstyle{\sum_j}} \mathbf{n}_{ij} \cdot  \mathbf{h}_j.
\label{eq:peffi2}
\end{equation}

\subsection{Stress as a map between networks}

\begin{figure}
\begin{center}
\includegraphics[height=2.5in]{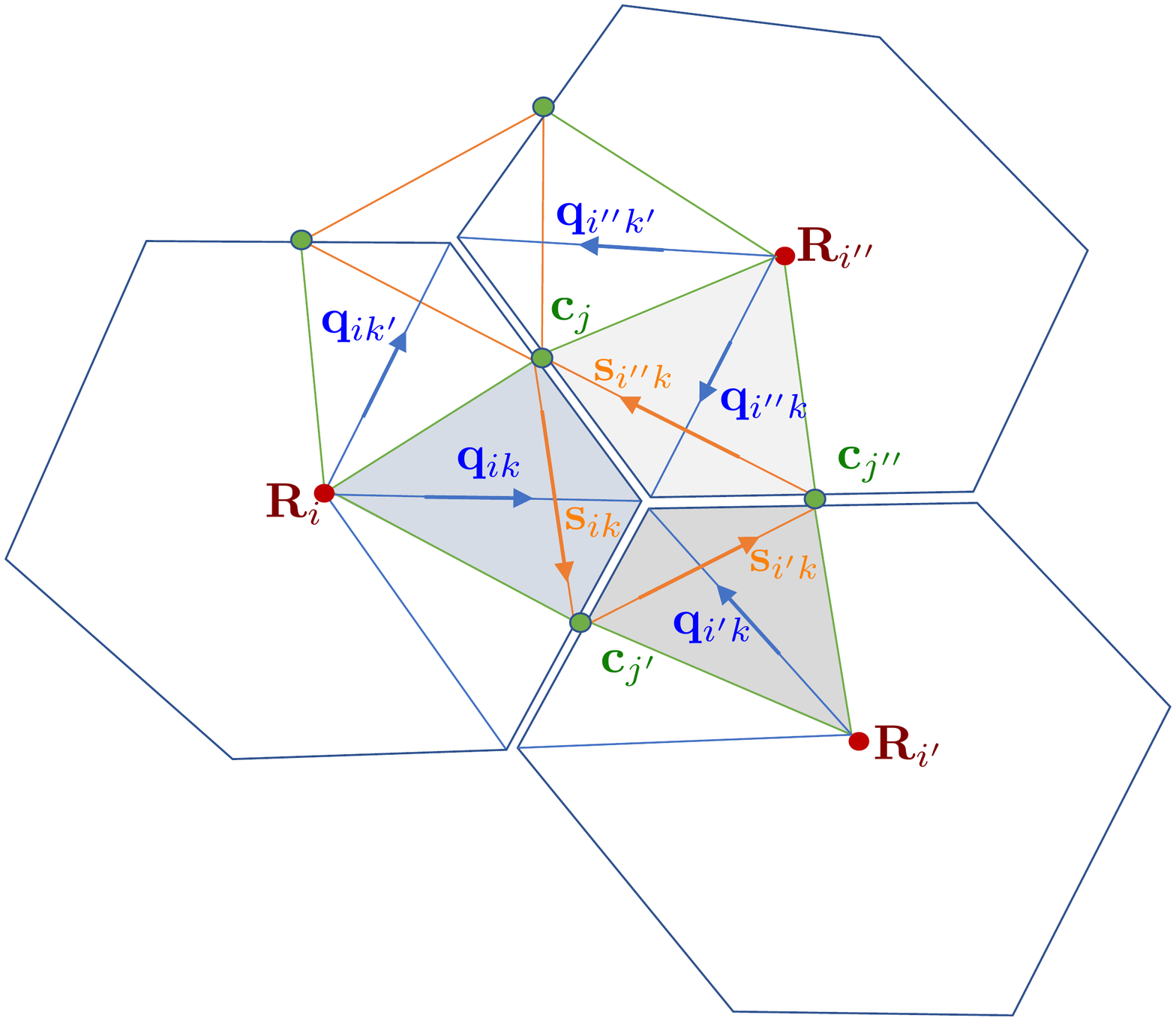}
\end{center}
\caption{Three cells, labelled $i$, $i'$ and $i''$, sharing vertex $k$ and edges $j$, $j'$ and $j''$ (taken anticlockwise).  In cell $i$, spoke $\mathbf{q}_{ik}$ connects cell centre $\mathbf{R}_i$ to vertex $\mathbf{r}_k$, intersecting the link $\mathbf{s}_{ik}$ between neighbouring edge centroids $\mathbf{c}_j$ and $\mathbf{c}_{j'}$.  Kite $ik$ is spanned by $\mathbf{q}_{ik}$ and $\mathbf{s}_{ik}$; the three kites neighbouring vertex $k$, that together form a tristar, are shaded.   The Airy stress function $\psi_{ik}$ (see below) is defined on kites.  Jumps in $\psi_{ik}$ between neighbouring kites in the same cell, sharing vertex $\mathbf{c}_j$, are defined by the projection of the force potential $\mathbf{h}_j$ on the cell edge $\mathbf{t}_j$.  Jumps in $\psi_{ik}$ between neighbouring kites in the same tristar, sharing vertex $\mathbf{c}_j$, are defined by the projection of  $\mathbf{h}_j$ on the link $\mathbf{T}_j$ between cell centres (not shown) that intersects edge $\mathbf{t}_j$.}
\label{fig:airy}
\end{figure}

We can compare (\ref{eq:kites}), which constructs cell area from kite areas, to stress written as (\ref{eq:sigA}), 
suggesting that stress can also be interpreted as a mapping between $\mathbf{s}_{ik}$ and $-\boldsymbol{\varepsilon} \mathbf{f}_{ik}$.  An explicit construction for such a map was provided in \cite{ball2002}.  The mapping $\mathsf{M}_k$ between vertices $\mathbf{c}_j$, $\mathbf{c}_{j'}$, $\mathbf{c}_{j''}$, running anticlockwise around a triangle surrounding vertex $k$ (Fig.~\ref{fig:airy}), to $\mathbf{h}_{j}$, $\mathbf{h}_{j'}$, $\mathbf{h}_{j''}$, also ordered anticlockwise, is
\begin{align}
2a_k\mathsf{M}_k=&
\mathbf{h}_j \left[\boldsymbol{\varepsilon} (\mathbf{c}_{j'}-\mathbf{c}_{j''})\right]^T+
\mathbf{h}_{j'} \left[\boldsymbol{\varepsilon} (\mathbf{c}_{j''}-\mathbf{c}_j)\right]^T+
\mathbf{h}_{j''} \left[\boldsymbol{\varepsilon} (\mathbf{c}_{j}-\mathbf{c}_{j'})\right]^T \nonumber \\
=&\left[ \mathbf{h}_j\otimes \mathbf{s}_{i'k}+\mathbf{h}_{j'}\otimes \mathbf{s}_{i''k}+\mathbf{h}_{j''}\otimes \mathbf{s}_{ik}\right] \boldsymbol{\varepsilon},
\label{eq:mkstrX}
\end{align}
where the triangle area $a_k$ satisfies $2a_k=(\boldsymbol{\varepsilon}\mathbf{s}_{ik})^T \mathbf{s}_{i''k} = (\boldsymbol{\varepsilon}\mathbf{s}_{i'k})^T \mathbf{s}_{ik} = (\boldsymbol{\varepsilon}\mathbf{s}_{i''k})^T \mathbf{s}_{i'k}$
and $\mathbf{s}_{ik}=\mathbf{c}_{j'}-\mathbf{c}_j$, $-\boldsymbol{\varepsilon}\mathbf{f}_{ik}=\mathbf{h}_{j'}-\mathbf{h}_j$, etc.  
The action of the map is demonstrated via
\begin{equation}
\mathsf{M}_k\mathbf{s}_{ik}=\frac{1}{2a_k} \left[ 
\mathbf{h}_j \left[-\boldsymbol{\varepsilon} \mathbf{s}_{i'k}\right]^T+
\mathbf{h}_{j'} \left[-\boldsymbol{\varepsilon} \mathbf{s}_{i''k}\right]^T+
\mathbf{h}_{j''} \left[-\boldsymbol{\varepsilon} \mathbf{s}_{ik}\right]^T
\right]\mathbf{s}_{ik}=\mathbf{h}_{j'}-\mathbf{h}_j=-\boldsymbol{\varepsilon} \mathbf{f}_{ik}.
\label{eq:mkstr}
\end{equation}


With the map defined, we can write $\mathbf{f}_{ik}=\boldsymbol{\varepsilon} \mathsf{M}_k\mathbf{s}_{ik}$.  Then from (\ref{eq:sigA}),
\begin{equation}
A_i \boldsymbol{\sigma}_i={\textstyle{\sum_k}} {C}_{ik} \mathbf{q}_{ik} \mathbf{f}_{ik}^T
=-{\textstyle{\sum_k}}{C}_{ik} \mathbf{q}_{ik} (\boldsymbol{\varepsilon}^T \mathsf{M}_k \mathbf{s}_{ik})^T= -{\textstyle{\sum_k}} {C}_{ik} (\mathbf{q}_{ik} \otimes \mathbf{s}_{ik}) \mathsf{M}_k^T \boldsymbol{\varepsilon}.
\label{eq:sigk}
\end{equation}
This shows how the cell's stress is built from the kite shape tensor $\mathbf{q}_{ik}\otimes {\mathbf{s}}_{ik}$ in (\ref{eq:2chaintri}), weighted by contributions $-\mathsf{M}_k^T \boldsymbol{\varepsilon}$ from the cell's vertices.   Accordingly, the stress over a tristar built from the three kites surrounding vertex $k$ is
\begin{equation}
E_k \boldsymbol{\sigma}_k=-{ \textstyle{\sum_i} }{C}_{ik} (\mathbf{q}_{ik} \otimes \mathbf{s}_{ik}) \mathsf{M}_k^T \boldsymbol{\varepsilon}= -\left(\boldsymbol{\varepsilon} E_k+ \mathsf{F}_k\right)\mathsf{M}_k^T \boldsymbol{\varepsilon},
\label{eq:sigX}
\end{equation}
where the fabric tensor $\mathsf{F}_k$ is given by (\ref{eq:fabric}).   

The stress over the tristar at vertex $k$ can also be written in terms of spokes $\mathbf{q}_{ik}$ and vertex forces $\mathbf{f}_{ik}$.  Replacing the latter with the force potential $\mathbf{h}_j$ and reordering, we find that multiplying each $\mathbf{h}_j$ is the difference between neighbouring spokes, \hbox{i.e.} the straight link $\mathbf{T}_j$ between cell centres.  In general, this does not pass through $\mathbf{c}_j$, so contributing to non-zero $\mathsf{F}_k$.  
Explicitly, using (\ref{eq:fint}),
\begin{align}
E_k \boldsymbol{\sigma}_k&={\textstyle{\sum_i} }{C}_{ik} \mathbf{q}_{ik} \mathbf{f}_{ik}^T 
=\sum_{i,j} B_{ij} A_{jk} \mathbf{q}_{ik} \mathbf{h}_j^T \boldsymbol{\epsilon}_i  \nonumber
\\
&={\textstyle{\sum_j} } \left[ {\textstyle{\sum_{i} }} B_{ij}  (\mathbf{r}_{k}-\mathbf{R}_i)  \right] A_{jk}\mathbf{h}_j^T \boldsymbol{\epsilon}_i
={\textstyle{\sum_{j} }}  A_{jk} \left(\mathbf{T}_{j} \otimes \mathbf{h}_j\right) \boldsymbol{\epsilon}_k,
\label{eq:vstcu}
\end{align} 
where $\mathbf{T}_j=\sum_i B_{ij}\mathbf{R}_i$ and $\boldsymbol{\epsilon}_k$ is the orientation of the triangle surrounding vertex $k$.  (As explained in Appendix~\ref{sec:bo}, we assume that all cells have the same orientiation $\boldsymbol{\epsilon}_i$, and that $\boldsymbol{\epsilon}_k=-\boldsymbol{\epsilon}_i$ uniformly; with fixed $j$ and $k$, $B_{ij}+B_{i'j}=0$ for all options in Fig.~\ref{fig:pridu} below, so that the $\mathbf{r}_k$ terms cancel in (\ref{eq:vstcu}).)  
The outward normals to triangle $k$ are $\mathbf{N}_{jk}=-\boldsymbol{\epsilon}_k A_{jk}\mathbf{T}_j$.  It follows that, analogously to (\ref{eq:peffi2}), the isotropic component of tristar stress is
\begin{equation}
P_{\mathrm{eff},k}=\tfrac{1}{2}\mathrm{tr}(\boldsymbol{\sigma}_k)=-\frac{1}{2E_k} {\textstyle{\sum_j} } \mathbf{N}_{jk} \cdot \mathbf{h}_j.
\label{eq:peffk2}
\end{equation}





\subsection{Expressing stress in terms of the Airy stress function}

To enforce zero torque on cell $i$, given by (\ref{eq:cellstr1}), we define a discrete potential $\psi_{ik}$ (the discrete Airy stress function, {which assigns a scalar value to each kite of the monolayer}) satisfying 
\begin{subequations}
\begin{equation}
{\textstyle{\sum_k}} A_{jk}\psi_{ik} = \mathbf{h}_j\cdot\mathbf{t}_j
\end{equation}
for either of the cells neighbouring edge $j$, \hbox{i.e.} with $\overline{B}_{ij}=1$, which automatically satisfies $\sum_{j,k} B_{ij} A_{jk} \psi_{ik}=0$ (because $\mathsf{B}\mathsf{A}=\mathsf{0}$ --- see Appendix~\ref{sec:bo}).   Likewise, zero torque on tristar $k$ requires, from (\ref{eq:vstcu}), $\sum_j A_{jk} \mathbf{h}_j\cdot \mathbf{T}_j=0$, which we satisfy with potential $\psi_{ik}$ satisfying 
\begin{equation}
{\textstyle{\sum_i}} B_{ij} \psi_{ik}=\mathbf{h}_j \cdot \mathbf{T}_j
\end{equation}
\label{eq:psidiff}
\end{subequations}
for both pairs of kites bounding edge $j$ (\hbox{i.e.} with $\overline{A}_{jk}=1$).  Pursuing the analogy with planar elasticity, we seek to define $\mathbf{h}_j$ as a discrete curl of $\psi_{ik}$, here evaluated over the spokes $\mathbf{q}_{ik}$ of the four kites surrounding $\mathbf{c}_j$ (\hbox{e.g.} the path $\mathbf{q}_{ik}-\mathbf{q}_{i''k}+\mathbf{q}_{i''k'}-\mathbf{q}_{ik'}$ in Fig.~\ref{fig:airy}) in order to recover the stress in terms of $\psi_{ik}$.

To illustrate the definition of $\psi_{ik}$, consider the three kites surrounding vertex $k$ (Fig.~\ref{fig:airy}), noting that links between neighbouring cells can be expressed in terms of spokes.  With $i$, $i'$, $i''$ and $j$, $j'$ and $j''$ ordered anticlockwise around vertex $k$, we require from (\ref{eq:psidiff}) that
\begin{subequations}
\begin{align}
\psi_{i'k}-\psi_{ik}&=(\mathbf{q}_{ik}-\mathbf{q}_{i'k})\cdot\mathbf{h}_{j'}, \\
\psi_{i''k}-\psi_{i'k}&=(\mathbf{q}_{i'k}-\mathbf{q}_{i''k})\cdot\mathbf{h}_{j''}, \\
\psi_{ik}-\psi_{i''k}&=(\mathbf{q}_{i''k}-\mathbf{q}_{ik})\cdot\mathbf{h}_{j}.
\end{align}
\label{eq:triloop}
\end{subequations}
Likewise, at neighbouring vertex $k'$, we require
\begin{equation}
\psi_{i''k'}-\psi_{ik'}=(\mathbf{q}_{ik'}-\mathbf{q}_{i''k'})\cdot\mathbf{h}_{j} 
=(\mathbf{q}_{ik}-\mathbf{q}_{i''k'})\cdot\mathbf{h}_{j}  =\psi_{i''k}-\psi_{ik}, 
\label{eq:qloop}
\end{equation}
showing that the jump in $\psi$ across edge $j$ is symmetric between neighbouring kites.  Accordingly, we can define  averages of $\psi_{ik}$ over neighbouring elements, $\phi_{ij}=\tfrac{1}{2}\sum_k \overline{A}_{jk}\psi_{ik}$ and $\theta_{jk}=\sum_i \tfrac{1}{2} \overline{B}_{ij}\psi_{ik}$, so that
\begin{subequations}
\begin{align}
\sum_{i,k}\tfrac{1}{2} \overline{A}_{jk}B_{ij}\psi_{ik}=&{ \textstyle{\sum_i} } B_{ij}\phi_{ij}\equiv {\textstyle{\sum_i} } B_{ij} \psi_{ik} \quad\mathrm{for~} k\mathrm{~such~that~} \overline{A}_{jk}=1, \\
\sum_{i,k}\tfrac{1}{2} \overline{B}_{ij}A_{jk}\psi_{ik}=&{\textstyle{\sum_k}} A_{jk}\theta_{jk}\equiv {\textstyle{\sum_k}} A_{jk} \psi_{ik} \quad\mathrm{for~}i\mathrm{~such~that~} \overline{B}_{ij}=1.
\end{align}
\label{eq:thetaphi}
\end{subequations}

We now express $\mathbf{h}_j$ in terms of the $\psi_{ik}$ in the four neighbouring kites (\hbox{i.e.} inverting expressions such as (\ref{eq:triloop})), using  the network of spokes.   Equation (\ref{eq:b4}) (Appendix~\ref{sec:kites}) demonstrates how a vector $\mathbf{g}$ can be constructed as a discrete curl of a potential defined across a diamond spanned by non-parallel vectors $\mathbf{a}$ and $\mathbf{b}$.   Assuming there are two jumps in potential, when crossing $\mathbf{a}$ and $\mathbf{b}$ respectively, with the jumps proportional to $\mathbf{g}\cdot\mathbf{a}$ and $\mathbf{g}\cdot\mathbf{b}$ but not a linear combination of the two (as is the case for $\mathbf{h}_j$, $\mathbf{t}_j$ and $\mathbf{T}_j$ in (\ref{eq:psidiff})), 
it is necessary for $\mathbf{a}\cdot\mathbf{b}=0$.  We therefore require
\begin{equation}
\mathbf{t}_j\cdot\mathbf{T}_j=0,
\label{eq:orthog}
\end{equation} 
\hbox{i.e.} each link between adjacent cell centres must intersect the corresponding cell edge orthogonally. 
Eq.~(\ref{eq:orthog}) is therefore necessary for both $\boldsymbol{\sigma}_i$ and $\boldsymbol{\sigma}_k$ to be symmetric (equivalently, for each to be expressed in terms of $\psi_{ik}$).  For the jumps in $\mathbf{h}_j\cdot \mathbf{t}_j$ and $\mathbf{h}_j\cdot \mathbf{T}_j$ to align appropriately with $\mathbf{t}_j$ and $\mathbf{T}_j$, a rotation and rescaling are necessary as in (\ref{eq:curot}), to give 
\begin{equation}
\mathbf{h}_j=\frac{1}{t_j T_j} \sum_{i,k} \boldsymbol{\epsilon}_i B_{ij} A_{jk} \psi_{ik} \mathbf{q}_{ik}.
\label{eq:hjcurl}
\end{equation}

Given (\ref{eq:orthog}), we can also express the force potential directly in terms of edges and links as 
\begin{equation}
\mathbf{h}_j\equiv \frac{(\mathbf{h}_j\cdot\mathbf{t}_j)\mathbf{t}_j}{t_j^2} + \frac{(\mathbf{h}_j\cdot\mathbf{T}_j)\mathbf{T}_j}{T_j^2}=
\sum_{i,k} \tfrac{1}{2}\overline{B_{ij}} A_{jk} \psi_{ik} \frac{\mathbf{t}_j}{t_j^2} + \sum_{i,k} \tfrac{1}{2} \overline{A}_{jk} B_{ij}\psi_{ik}\frac{\mathbf{T}_j}{T_j^2}.
\label{eq:hj}
\end{equation}
Recalling that $\boldsymbol{\epsilon}_i\mathbf{t}_j$ defines a normal to edge $j$ relative to cell $i$,  we see that
\begin{equation}
(\boldsymbol{\epsilon}_i\mathbf{t}_j)\cdot \mathbf{h}_j=
 \sum_{i,k} \tfrac{1}{2} \overline{A}_{jk} B_{ij}\psi_{ik}\frac{(\boldsymbol{\epsilon}_i\mathbf{t}_j)\cdot \mathbf{T}_j}{T_j^2} =
 \sum_{i,k} \tfrac{1}{2} \overline{A}_{jk} B_{ij}\psi_{ik}({t_j}/{T_j}),
\end{equation}
noting that, for all four cases in Fig.~\ref{fig:pridu} below, $(\boldsymbol{\epsilon}_i\mathbf{t}_j)\cdot \mathbf{T}_j=t_jT_j$.  Thus from (\ref{eq:peffi}),  and using (\ref{eq:thetaphi}), we obtain an alternative to (\ref{eq:peffi2})
\begin{equation}
P_{\mathrm{eff},i}=\frac{1}{2A_i} {\textstyle{\sum_j} } B_{ij} (\boldsymbol{\epsilon}_i \mathrm{t}_j)\cdot \mathbf{h}_j=
\frac{1}{4A_i}\sum_{i',j,k} B_{ij} \frac{t_j}{T_j}  B_{i'j}\overline{A}_{jk}\psi_{i'k} \equiv
\frac{1}{2A_i}\sum_{i',j} B_{ij} \frac{t_j}{T_j}  B_{i'j}\phi_{i'j}.
\label{eq:peffilap}
\end{equation}
As might be expected from classical elasticity, the isotropic component of the stress is given as a discrete Laplacian (over the primary network) of the Airy stress function, involving (for an interior cell) $3Z_i$ kites and $2Z_i$ independent values of $\psi_{ik}$.  Likewise, noting that $(\boldsymbol{\epsilon}_k\mathbf{t}_j)\cdot \mathbf{T}_j=t_jT_j$, the isotropic stress over tristars is given by a Laplacian over the dual network involving (for an interior cell) 9 kites and 6 independent values of $\psi_{ik}$, namely
\begin{equation}
P_{\mathrm{eff},k}=\frac{1}{4E_k}\sum_{i,j,k'} A_{jk}\frac{T_j}{t_j}  A_{jk'}\overline{B}_{ij}\psi_{ik'}\equiv 
\frac{1}{2E_k}\sum_{j,k'} A_{jk}\frac{T_j}{t_j}  A_{jk'}\theta_{jk'},
\label{eq:peffk}
\end{equation}
providing an alternative to (\ref{eq:peffk2}).

Finally, we can write the tristar stress in terms of links and edges using (\ref{eq:vstcu}, \ref{eq:hj}) as 
\begin{equation}
E_k \boldsymbol{\sigma}_k={ \textstyle{\sum_j} } A_{jk}\mathbf{T}_j\otimes \left[
\sum_{i,k'} \tfrac{1}{2}\overline{B_{ij}} A_{jk'} \psi_{ik'} \frac{\mathbf{t}_j}{t_j^2} + \sum_{i,k'} \tfrac{1}{2} \overline{A}_{jk'} B_{ij}\psi_{ik'}\frac{\mathbf{T}_j}{T_j^2} \right]\boldsymbol{\epsilon}_k.
\label{eq:hja}
\end{equation}
Now $\mathbf{t}_j^T\boldsymbol{\epsilon}_k=\mathbf{T}_j^T (t_j/T_j)$ and $\mathbf{T}_j^T \boldsymbol{\epsilon}_k=-\mathbf{t}_j^T (T_j/t_j)$ in each of the four cases illustrated in Fig.~\ref{fig:pridu} below.  Thus, making use of (\ref{eq:thetaphi}) and (\ref{eq:peffk}),
\begin{equation}
E_k \boldsymbol{\sigma}_k={\textstyle{\sum_j} } \frac{A_{jk}}{t_j T_j} \mathbf{T}_j\otimes \left[
{\textstyle{\sum_{k'} }} A_{jk'} \theta_{jk'} {\mathbf{T}_j} 
-{\textstyle{ \sum_{i} }} B_{ij}\psi_{ik} {\mathbf{t}_j} \right]=E_k P_{\mathrm{eff},k}\mathsf{I}-
\sum_{i,j} B_{ij} \frac{\mathbf{T}_j\otimes  \mathbf{t}_j}{T_j t_j}   A_{jk}\psi_{ik},
\label{eq:hjb}
\end{equation}
showing how the shear stress is captured by differences in the $\psi_{ik}$ field between neighbouring kites intersecting the tristar.  Likewise, using the identities $\mathbf{t}_j^T\boldsymbol{\epsilon}_i=-\mathbf{T}_j^T (t_j/T_j)$ and $\mathbf{T}_j^T \boldsymbol{\epsilon}_i=\mathbf{t}_j^T (T_j/t_j)$, we find
\begin{equation}
A_i\boldsymbol{\sigma}_i=A_i P_{\mathrm{eff},i} \mathsf{I}-\sum_{j,k} B_{ij} \frac{\mathbf{t}_j \otimes \mathbf{T}_j }{t_jT_j} A_{jk} \psi_{ik}.
\label{eq:gi}
\end{equation}
In cell $i$, the final sum in (\ref{eq:gi}) allocates a scalar ($v_{ij}\equiv \sum_k B_{ij} A_{jk} \psi_{ik}$, the pairwise difference in $\psi_{ik}$ values taken in the same sense as the orientation of cell $i$) to each edge and then sums the outer products of the unit tangent and the inward unit normal, weighted by $v_{ij}$ and taken anticlockwise, such that $\sum_j v_{ij}=0$.

It is not immediately obvious that the stress tensors in (\ref{eq:hj}, \ref{eq:gi}) are still symmetric.  However writing the outer product of unit vectors as $\hat{\mathbf{t}}_j\otimes\hat{\mathbf{T}}_j= (\cos\alpha_j,\sin\alpha_j)^T(-\sin\alpha_j ,\cos\alpha_j)$ (when $\boldsymbol{\epsilon}_i=-\boldsymbol{\varepsilon}$), where $\alpha_j$ is the orientation of edge $j$ with respect to a fixed axis, then the final sum in (\ref{eq:gi}) is (for $\boldsymbol{\epsilon}_i=\pm \boldsymbol{\varepsilon}$)
\begin{equation}
\mathsf{D}_i\equiv {\textstyle{\sum_j} } v_{ij}\hat{\mathbf{t}}_j\otimes \hat{\mathbf{T}}_j = \boldsymbol{\varepsilon} \boldsymbol{\epsilon}_i { \textstyle{\sum_j} } v_{ij}\left( 
\begin{matrix} - \cos\alpha_j \sin\alpha_j & \cos^2 \alpha_j \\ -\sin^2 \alpha_j & \cos\alpha_j \sin\alpha_j \end{matrix} 
\right).
\end{equation}
This is symmetric because $\sum_j v_{ij}\cos^2 \alpha_j = \sum_j v_{ij} (1-\sin^2 \alpha_j)=-\sum_j v_{ij} \sin^2 \alpha_j$.  Thus
\begin{equation}
\mathsf{D}_i=\tfrac{1}{2}\left (\mathsf{D}_i+\mathsf{D}_i^T \right) = \tfrac{1}{2} \boldsymbol{\varepsilon} \boldsymbol{\epsilon}_i { \textstyle{\sum_j} }v_{ij}\left( 
\begin{matrix} - \sin 2 \alpha_j & \cos2 \alpha_j \\ \cos 2\alpha_j  & \sin2\alpha_j \end{matrix} 
\right),
\label{eq:devi}
\end{equation}
which we will make use of shortly.

\subsection{Tristar stress and the fabric tensor}
\label{sec:tristar}

We now reconcile the two expressions for $\boldsymbol{\sigma}_k$ in (\ref{eq:sigX}) and (\ref{eq:vstcu}).  First, consider the condition
\begin{equation}
{A}_{jk} \frac{\mathbf{T}_j\boldsymbol{\epsilon}_k}{E_k} =\overline{A}_{jk} {\textstyle{\sum_i}}  (C_{ik}-\overline{B}_{ij})\frac{ \mathbf{s}_{ik} \boldsymbol{\varepsilon}}{2a_k}  .
\label{eq:linksim}
\end{equation}
Link $\mathbf{T}_j$ crosses edge $\mathbf{t}_j$, bounded by two vertices, each surrounded by a triangle of vectors $\mathbf{s}_{ik}$ of the edge centroid network having area $a_k$; two such triangles are illustrated in Fig.~\ref{fig:airy}.  Depending on the chosen orientation of cells and edges, (\ref{eq:linksim}) implies that link $\mathbf{T}_j$ is parallel (or antiparallel) to the furthest edge of each triangle (such as $\mathbf{s}_{i'k}$ in Fig.~\ref{fig:airy}), with the magnitude of $\mathbf{T}_j$ relative to the edge $\mathbf{s}_{ik}$ given by the ratio $E_k/2a_k$.   In other words, (\ref{eq:linksim}) implies that each vertex bounding $\mathbf{t}_j$ lies at the orthocentre of the triangle of edge-centroid-links surrounding each vertex.  

Direct substitution of (\ref{eq:linksim}) into (\ref{eq:vstcu}), giving $\boldsymbol{\sigma}_k$ as an outer product of links with the force potential, recovers $-\boldsymbol{\varepsilon} \mathsf{M}^T_k\boldsymbol{\varepsilon}$ with $\mathsf{M}_k$ defined in (\ref{eq:mkstrX}), given as an outer product of edge-centroid-links with the force potential.  Thus (\ref{eq:linksim}) is equivalent to the condition
\begin{equation}
\boldsymbol{\sigma}_k=-\boldsymbol{\varepsilon} \mathsf{M}_k^T \boldsymbol{\varepsilon}.
\label{eq:strmap}
\end{equation}
Symmetry of $\boldsymbol{\sigma}_k$ is ensured by the existence of the Airy stress function and the orthogonality condition (\ref{eq:orthog}); (\ref{eq:linksim}) extends this symmetry to $\mathsf{M}_k$.  Furthermore, (\ref{eq:sigX}) then implies that $\mathsf{F}_k \mathsf{M}_k=\mathsf{0}$, while (\ref{eq:strmap}) gives $\mathsf{M}_k=-\boldsymbol{\varepsilon}\boldsymbol{\sigma}_k  \boldsymbol{\varepsilon}$, yielding the \textit{stress-geometry condition}  \cite{ball2002, blumenfeld2003, degiuli2014}
\begin{equation}
\mathsf{F}_k\boldsymbol{\varepsilon} \boldsymbol{\sigma}_k=\mathsf{0}.
\label{eq:fes}
\end{equation}  
The role of stress as a mapping between networks is also evident via 
$\mathbf{f}_{ik}=\boldsymbol{\sigma}_k\boldsymbol{\varepsilon} \mathbf{s}_{ik}$, showing how a force balance can be turned into a divergence of stress (via (\ref{eq:foreq})).   The mapping can also be used to show that the area of the triangle $k$ in the force network is $\mathrm{det}(\boldsymbol{\sigma}_k)$ times that in the edge centroid network (Appendix~\ref{sec:areamap}).

We can use (\ref{eq:fes}) to infer stress orientation in the neighbourhood of a vertex. As long as $\mathrm{det}(\boldsymbol{\sigma}_k)\neq 0$, we can write $\boldsymbol{\sigma}_k$ in terms of its principal axes and eigenvalues as $\boldsymbol{\sigma}_k=\sigma_{k1}\mathbf{e}_{k1}\otimes \mathbf{e}_{k1} + \sigma_{k2}\mathbf{e}_{k2}\otimes \mathbf{e}_{k2}$, where $\mathbf{e}_{k1}\cdot\mathbf{e}_{k2}=0$.   Likewise, as long as $\mathrm{det} (\mathsf{F}_k) \neq 0$, we may express it in terms of its principal axes as $\mathsf{F}_k=F_{k1} \mathbf{f}_{k1}\otimes\mathbf{f}_{k1}+F_{k2} \mathbf{f}_{k2}\otimes\mathbf{f}_{k2}$ where $\mathbf{f}_{k1}\cdot\mathbf{f}_{k2}=0$. Then (\ref{eq:fes}) implies 
\begin{multline}
F_{k1} \sigma_{k1}(\mathbf{f}_{k1}^T\boldsymbol{\varepsilon}\mathbf{e}_1)  \mathbf{f}_{k1}  \otimes \mathbf{e}_1 + 
F_{k1} \sigma_{k2}(\mathbf{f}_{k1}^T\boldsymbol{\varepsilon}\mathbf{e}_2)  \mathbf{f}_{k1}  \otimes \mathbf{e}_2 + 
F_{k2} \sigma_{k1}(\mathbf{f}_{k2}^T\boldsymbol{\varepsilon}\mathbf{e}_1)  \mathbf{f}_{k2}  \otimes \mathbf{e}_1 \\ + 
F_{k2} \sigma_{k2}(\mathbf{f}_{k2}^T\boldsymbol{\varepsilon}\mathbf{e}_2)  \mathbf{f}_{k2}  \otimes \mathbf{e}_2 
=\mathsf{0}. 
\end{multline} 
This will be satisfied when the orthogonal axes of each tensor align, so that the cross products vanish.  The fabric tensor therefore provides a direct mechanism for inferring stress orientation in the neighbourhood of vertices, except when there is sufficient symmetry for the fabric tensor to vanish.


\subsection{Relating cell centres and cell vertices}

We have not yet specified how cell centres $\mathbf{R}_i$ might be related to cell vertices $\mathbf{r}_k$, so that conditions (\ref{eq:orthog}) and (\ref{eq:linksim}) may be satisfied.  The orthogonality condition (\ref{eq:orthog}) applies to all border and internal edges: the links are $\sum_i B_{ij}^{bb}\mathbf{R}^b_i$, $\sum_i(B_{ij}^{bi} \mathbf{R}_i^b+B^{ii}_{ij}\mathbf{R}_i^i)$ and the edges $\sum_k (A_{jk}^{bp}\mathbf{r}^p_k+ A_{jk}^{bi} \mathbf{r}_k^i)$, $\sum_k A^{ii}_{jk}\mathbf{r}_k^i$.  Then (\ref{eq:orthog}) requires, for $N_e-N_p$ border and internal edges
\begin{equation}
\sum_{i,k} B_{ij}^{bb}\mathbf{R}^b_i \cdot  (A_{jk}^{bp}\mathbf{r}^p_k+ A_{jk}^{bi} \mathbf{r}_k^i) =0,\quad
\sum_{i,k} (B_{ij}^{bi} \mathbf{R}_i^b+B^{ii}_{ij}\mathbf{R}_i^i) \cdot A^{ii}_{jk}\mathbf{r}_k^i =0.
\label{eq:centroids}
\end{equation}
The $\mathbf{R}_i$ correspond to $2N_c$ scalar quantities.  Given a set of vertices, for $2N_c>N_e-N_p$ the system is underconstrained and one expects to find many possible cell centre locations  for which (\ref{eq:orthog}) is satisfied (\hbox{i.e.}, for a small number of cells, it is easy to construct a triangulation of cell centres for which links are orthogonal to edges).   However for larger monolayers, with $2N_c<N_e-N_p$ (anticipating that $N_e\sim 3N_c$ and $N_p\propto \sqrt{N_c}$ for $N_c\gg 1$), then (\ref{eq:centroids}) becomes overconstrained.   In other words, a set of vertex locations emerging from a simulation that does not impose a moment balance cannot be expected, in general, to admit a triangulation satisfying (\ref{eq:centroids}).   Similarly, Fig.~\ref{fig:ex}(d) shows how (\ref{eq:centroids}) is violated when cell centres are chosen to be cell centroids, satisfying $\mathbf{R}_i=\tilde{\mathbf{R}}_i$, where
\begin{equation}
\tilde{\mathbf{R}}_i\equiv \frac{1}{Z_k} {\textstyle{\sum_k }} C_{ik}\mathbf{r}_k.
 \label{eq:centroid2}
\end{equation}

\begin{figure}
\begin{center}
\includegraphics[height=3in]{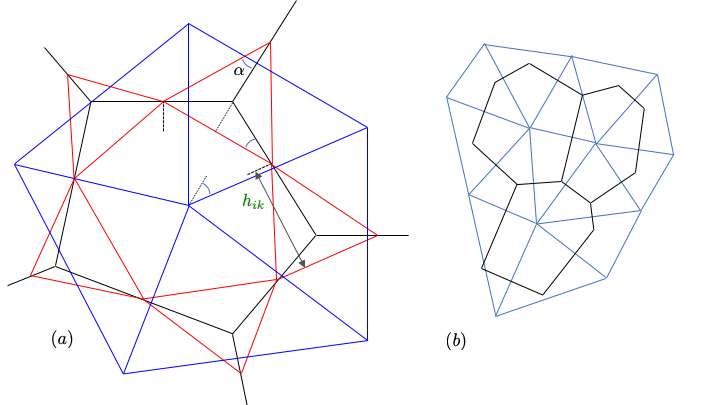}
\end{center}
\caption{(a) Geometric constructions demanded by torque balance.  Red triangles connect adjacent edge centroids.  Each cell vertex is at its orthocentre, \hbox{i.e.} the (black) cell edge passing through a vertex of a red triangle is orthogonal to the opposite side of the red triangle.   The links between adjacent cell centres (blue) are orthogonal to the cell edges they intersect.  Thus each red triangle surrounding a vertex is similar to the blue triangle surrounding the same vertex (opposite edges are parallel), differing by a rotation of $\pi$ and uniform scaling.  Dashed lines are orthogonal to cell edges and dotted lines are orthogonal to triangle edges.  The angles marked with arcs are equal.  $h_{ik}$ indicates the altitude of the edge-centroid triangle at vertex $k$. 
{(b) An illustration of three cells (black) satisfying orthocentric constraints and a corresponding network of cell centres (blue), constructed using an algorithm described in Appendix~\ref{sec:dual}.}}
\label{fig:selfsim}
\end{figure}

The constraint (\ref{eq:linksim}) can be interpreted from a cell-based perspective (in which cell vertices $\mathbf{r}$ define cell centres $\mathbf{R}$), as 
illustrated in Fig.~\ref{fig:selfsim}(a).  The vertices $\cup_k C_{ik} \mathbf{r}_k$ of cell $i$ define its edges $\cup_j \overline{B}_{ij}\mathbf{t}_j$, edge centroids $\cup_j \overline{B}_{ij}\mathbf{c}_j$ and the internal links between them $\cup_k C_{ik}\mathbf{s}_{ik}$.  Eq.~(\ref{eq:linksim}) requires that the edge radiating outwards from cell $i$ at vertex $k$ must be orthogonal to $\mathbf{s}_{ik}$; the triangle of links $\cup_i C_{ik} \mathbf{s}_{ik}$ around vertex $k$ is then fully defined since $\mathbf{r}_k$ is at its orthocentre.  This in turn specifies the centroid (and therefore length) of each edge radiating from the cell.  A triangulation of the plane is then constructed (in principle) via each triangle of the edge-centroid network at each vertex being rotated by $\pi$ and expanded by a factor $1/\lambda_k$, say, under the constraint (\ref{eq:linksim}), to cover the plane (Fig.~\ref{fig:selfsim}a), with $\mathbf{R}_i$ the common vertex of all triangles covering cell $i$.   
Noting that $2a_k$ is the product of the base and the height of the edge-centroid triangle at vertex $k$, link $j$ should have length $T_j$ satisfying (from (\ref{eq:linksim}))
\begin{equation}
\overline{B}_{ij} \overline{A}_{jk}(T_j- E_k/h_{ik})=0,
\label{eq:areasc}
\end{equation}
where $h_{ik}$ is the altitude of the triangle at vertex $k$ (Fig.~\ref{fig:selfsim}a).  Given that the area $U_k$ of the triangle spanned by $\cup_i C_{ik}\mathbf{R}_i$ maps to $a_k$ via $a_k=\lambda_k^2 U_k$, with altitudes related by $h_{ik}=\lambda_k H_{ik}$, (\ref{eq:areasc}) implies $E_k=T_jh_{ik}=2 \lambda_k U_k=2a_k/\lambda_k$, implying 
\begin{equation}
E_k=2(a_k U_k)^{1/2},
\label{eq:tristararea}
\end{equation} 
a result that can be verfied by (\ref{eq:2chaintri}).  

The covering of the plane by rotated and expanded triangles requires the internal angles at the vertices of the triangles to sum to $2\pi$ where they meet at $\mathbf{R}_i$.  (To demonstrate that this is feasible, consider the closed triangles linking edge centroids around the vertices of cell $i$.  The outermost vertex of each triangle is bisected by an edge separating two neighbouring cells.  The resulting angle $\alpha$, marked in Fig.~\ref{fig:selfsim}(a) appears also within another vertex of the triangle, indicating that $\pi/2-\alpha$ contributes to the internal angle of the polygon of links $\mathsf{s}_{ik}$ within cell $i$.  Since all such internal angles sum to $(Z_i-2)\pi$, it follows that all angles $\alpha$ sum to $2\pi$.)  The covering also requires consistent scaling between neighbouring triangles. { We show in  Appendix~\ref{sec:dual} how this condition is satisfied (up to a translation and uniform scaling of the triangulation), and how such a covering may extend to the whole monolayer.  Fig.~\ref{fig:selfsim}(b) illustrates such a covering for three cells, satisfying the orthocentric property (\ref{eq:orthog}, \ref{eq:linksim}).}


In summary, the zero-net-torque constraints described above, { specifically the requirement that each internal vertex lies at the orthocentre of the triangle formed by its neighbours (or equivalently its neighbouring edge centroids), can be used to define} a self-consistent set of cell centre locations.  


\section{Constitutive modelling}
\label{sec:con}

So far, we have assumed that the mechanical load applied to (or generated by) a cell can be approximated by forces applied at its vertices, without specifying how these might be related to the size and shape of the cell.  Commonly, cell $i$, with area $A_i$ and perimeter $L_i$, is assumed to have mechanical energy $U_i=\mathcal{U}(A_i, L_i)$, so that cells have identical mechanical properties but distinct shapes and sizes.   $\mathcal{U}$ typically includes a quadratic area-dependent term penalising departures from a reference area, that measures the resistance of the cytoplasm to expansion or contraction, and a quadratic perimeter-dependent term that penalises departures from a reference length, capturing the resistance to stretching of the cell cortex as may take place under shear.   These contributions define a pressure $\mathcal{P}_i$ and a tension $\mathcal{T}_i$ for each cell via
$\mathcal{P}_i\equiv {\partial U_i}/{\partial A_i}$, $\mathcal{T}_i\equiv {\partial U_i}/{\partial L_i}$.
Equations (\ref{eq:perim}) and (\ref{eq:area}), showing how the length and perimeter of cell $i$ change when vertex $k$ moves, can then be used to evaluate $\mathbf{f}_{ik}=\delta U_i/\delta \mathbf{r}_k$, the first variation of the energy of cell $i$ with respect to a small displacement of vertex $\mathbf{r}_k$.  This determines the restoring force at $\mathbf{r}_k$ due to cell $i$ as
\begin{equation}
\mathbf{f}_{ik}= {\textstyle{\sum_{j} }} \left[ \tfrac{1}{2} \mathcal{P}_i\boldsymbol{\epsilon}_i  B_{ij}  \mathbf{t}_j  \overline{A}_{jk}  +  \mathcal{T}_i  {\overline{B}}_{ij} \hat{\mathbf{t}}_j A_{jk}\right].
\label{eq:forver}
\end{equation}
We can evaluate the cell stress using (\ref{eq:a2}) and (\ref{eq:forver}), noting that $C_{ik}$ becomes redundant, as
\begin{subequations}
\begin{align}
A_i\boldsymbol{\sigma}_i=&\sum_{j,k} 
 \left[ \tfrac{1}{2} \mathcal{P}_i\boldsymbol{\epsilon}_i  B_{ij}  \mathbf{r}_k\otimes\mathbf{t}_j  \overline{A}_{jk}  +  \mathcal{T}_i  {\overline{B}}_{ij} \mathbf{r}_k\otimes \hat{\mathbf{t}}_j A_{jk}\right] \\
=& {\textstyle{ \sum_{j} }}
 \left[  \mathcal{P}_i \boldsymbol{\epsilon}_i  B_{ij}  \mathbf{c}_j\otimes\mathbf{t}_j   +  \mathcal{T}_i  {\overline{B}}_{ij} \mathbf{t}_j\otimes \hat{\mathbf{t}}_j \right] \\
=&  A_i \mathcal{P}_i \mathsf{I}   +{\mathcal{T}_i L_i} \mathsf{Q}_i ,
 \end{align}
\label{eq:cellstress}
\end{subequations}
where $\mathsf{Q}_i\equiv {L_i}^{-1}  { \textstyle{\sum_j} } \overline{B}_{ij} t_j {\hat{\mathbf{t}}_j\otimes \hat{\mathbf{t}}_j}$,
a result consistent with prior studies (\hbox{e.g.} \cite{guirao2015, ishihara2012, ANB2018a, yang2017}).  These terms can be interpreted by noting that under an imposed uniform strain $\mathsf{E}$, $A_i$ changes by $A_i \mathsf{I}:\mathsf{E}\equiv A_i \mathrm{Tr}(\mathsf{E})$ and $L_i$ changes by $L_i \mathsf{Q}_i:\mathsf{E}$ \cite{ANB2018b}.  The cell structure tensor $\mathsf{Q}_i$ commutes with the cell shape tensor $\sum_k \mathbf{q}_{ik} \otimes \mathbf{q}_{ik}$, implying that the principal axes of stress and shape align at the cell level \cite{ANB2018a}.  The isotropic component of the stress in each cell shows that the effective pressure
\begin{equation}
P_{\mathrm{eff},i}\equiv \tfrac{1}{2}\mathrm{tr}(\boldsymbol{\sigma}_i) =\mathcal{P}_i+\frac{\mathcal{T}_iL_i}{2A_i}
\label{eq:peffi}
\end{equation}
has contributions from both the cell's interior and its periphery.    

Writing (\ref{eq:cellstress}) as $A_i\boldsymbol{\sigma}_i=A_i P_{\mathrm{eff},i} \mathsf{I}+\mathcal{T}_i L_i (\mathsf{Q}_i-\tfrac{1}{2} \mathsf{I})$, comparison with (\ref{eq:gi}) shows that the intracellular differences in $\psi_{ik}$ are proportional to $\mathcal{T}_i L_i$.  More specifically, writing $\hat{\mathbf{t}}_j=(\cos \alpha_j,\sin\alpha_j)^T$ with respect to some fixed Cartesian axes, 
\begin{equation}
\mathsf{Q}_i-\tfrac{1}{2}\mathsf{I} = \tfrac{1}{2} {\textstyle{\sum_j} } \overline{B}_{ij}\frac{t_j}{L_i} \left(\begin{matrix}
\cos 2 \alpha_j &  \sin 2\alpha_j \\ \sin 2 \alpha_j & - \cos 2 \alpha_j
\end{matrix}\right),
\end{equation}
recalling that $L_i=\sum_j \overline{B}_{ij} t_j$.  Comparison with (\ref{eq:devi}) then enables us to express the intracellular jump in Airy stress function between kites explicitly, as given in Appendix~\ref{sec:air}, revealing that the variation of Airy stress function within a cell is $\mathcal{T}_i {L}_i$ times a nonlinear but dimensionless measure of cell shape.   

{We note that, since $\mathcal{U}$ is defined in terms of areas and perimeters, it clearly satisfies material frame indifference.  Under a change in reference frame, in which $\mathbf{r}\mapsto \mathsf{Y}\mathbf{r}+\mathbf{Z}$, where $\mathsf{Y}$ is a rotation and $\mathbf{Z}$ a translation, it is straightforward to show that $\mathbf{t}\mapsto \mathsf{Y}\mathbf{t}$ (as do other vectors contributing to $\mathbf{f}$ in (\ref{eq:forver})), while the identity $\boldsymbol{\varepsilon}= Y\boldsymbol{\varepsilon} \mathsf{Y}^T$ ensures that other tensors, such as stress, transform under $\boldsymbol{\sigma}\mapsto \mathsf{Y} \boldsymbol{\sigma}\mathsf{Y}^T$, and hence are also frame indifferent.}


The constitutive model can also be extended to accommodate viscous dissipation, either within the cell itself or as a result of substrate drag.  In the former case, at time $t$, $\mathcal{P}_i$ in (\ref{eq:forver})  is replaced with with $\mathcal{P}_i+\gamma \mathrm{d}A_i/\mathrm{d}t$, and $\mathcal{T}_i$ with $\mathcal{T}_i+\mu \mathrm{d}\mathcal{L}_i/\mathrm{d}t$, corresponding to a dissipation rate $\Phi_i\equiv \gamma( \mathrm{d}A_i/\mathrm{d}t)^2+\mu (\mathrm{d}\mathcal{L}_i/\mathrm{d}t)^2$ in cell $i$ for positive parameters $\gamma$  and $\mu$, imposing that total dissipation $\Phi=\sum_i \Phi_i$ is minimized subject to $\Phi=-\mathrm{d}U/\mathrm{d}t$  \cite{ANB2018b}, where $U=\sum_i U_i$.   In the latter case (which is much more widely implemented in the literature), a drag force imposed on vertex $k$ by cell $i$ of magnitude $\tfrac{1}{3} \eta \mathrm{d}\mathbf{r}_k/\mathrm{d}t$ is added to $\mathbf{f}_{ik}$ at each internal vertex for some $\eta>0$, leading effectively to $N_v$ coupled ODEs for $\mathbf{r}_k(t)$ of the form 
\begin{equation}
\eta \frac{\mathrm{d} {\mathbf{r}}_k}{\mathrm{d}t}=- {\textstyle{\sum_i} } {C}_{ik} \mathbf{f}_{ik},
\label{eq:verdyn}
\end{equation}
with $\mathbf{f}_{ik}$ given by (\ref{eq:forver}).
Both instances lead to identical conclusions in terms of the structure of the force network and of the Airy stress function in (\ref{eq:hj}, \ref{eq:gi}), for example, but differences in detail once the stress is expressed in terms of pressures and tensions.   Crucially, however, (\ref{eq:verdyn}) alone is insufficient to ensure moments are balanced across the monolayer.

In summary, 
tracking variations of energy (and possibly dissipation rate) in terms of displacement of individual vertices (rather than in terms of strains, as in conventional elasticity), and imposing force balances alone, are insufficient in general to guarantee a torque balance.  Extra constraints must be imposed on the evolution of the total energy $U$ as it moves towards equilibrium.  Conditions (\ref{eq:orthog}, \ref{eq:linksim}) together suggest that a constrained energy minimization of the form
\begin{equation}
\min_{\mathbf{r}_k} \left\{U- {\textstyle{\sum_{i,j,k}}} \lambda_{ik} \overline{A}_{jk} \left(C_{ik}-\overline{B}_{ij} \right)\mathbf{t}_j \cdot \mathbf{s}_{ik} \right\}
\label{eq:opt}
\end{equation}
might be used, introducing Lagrange multipliers $\lambda_{ik}$ that ensure that each internal vertex lies at the orthocentre of the triangle formed by adjacent edge centroids (Fig.~\ref{fig:selfsim}a).  Conveniently, (\ref{eq:opt}) involves cell edges and links between edge centroids that can be directly expressed in terms of vertex locations.  {Following the construction in Appendix~\ref{sec:dual}, we can construct a dual network that identifies cell centres and links, up to a translation and scaling.  The degree of freedom in scaling is accommodated by jumps in the Airy stress function across cell edges, but otherwise there is no impact on representations of stress.  
}

\section{Discussion}
\label{sec:disc}

 The planar vertex model describes cells as a network of polygons that tile a region of the plane.  We have shown that a natural dual network is one that connects cell centres (suitably defined) via the mid-points of cell edges, forming tristars around each vertex (Fig.~\ref{fig:regionsA}).  To represent force balances geometrically, further subdivision of these networks is required, into the links $\mathbf{s}_{ik}$ between adjacent edge centroids and spokes $\mathbf{q}_{ik}$ within each cell.  The building blocks of the primal (cellular) and dual (tristar) networks are kites, defined by $\mathbf{q}_{ik}\otimes \mathbf{s}_{ik}$.  The antisymmetric part of this outer product gives the oriented area of the kite in cell $i$ neighbouring vertex $k$; the symmetric part characterises asymmetries in tristar shape via the fabric tensor $\mathsf{F}_k$, defined in (\ref{eq:fabric}). 
 
 Stress in two dimensions (in continuum linear elasticity) can be written as the curl of a vector potential, which itself can be written as the curl of a scalar (the Airy stress function).  In the present problem, we have shown that the Airy stress function $\psi_{ik}$ is defined on kites and curls are discrete: the vector force potential $\mathbf{h}_j$ on edge $j$ can be constructed as a curl of $\psi_{ik}$ taken over adjacent spokes $\mathbf{q}_{ik}$ via (\ref{eq:hjcurl}), while cell stress $\boldsymbol{\sigma}_i$ is a curl of $\mathbf{h}_j$ taken around cell edges $\mathbf{t}_j$ via (\ref{eq:cellstr2}); likewise, tristar stress $\boldsymbol{\sigma}_k$ is a curl of $\mathbf{h}_j$ around links $\mathbf{T}_j$ between adjacent cell centres via (\ref{eq:vstcu}).   Jumps in $\psi_{ik}$ between neighbouring kites capture the projection of $\mathbf{h}_j$ onto $\mathbf{t}_j$ or $\mathbf{T}_j$: jumps across cell edges contribute to the isotropic stress, and jumps within cells across links contribute to shear stress.  We find that $\mathbf{t}_j\cdot\mathbf{T}_j=0$ is a necessary condition for $\mathbf{h}_j$ to be defined as a discrete curl of a potential having the appropriate jumps; equivalently, it is a necessary condition for a torque balance on cells and tristars.  However $\mathbf{t}_j$ and $\mathbf{T}_j$ need not intersect at edge or link centroids, and so networks differ in general from a classical {(or radical)} Voronoi construction  \cite{redenbach2009}.   We also identified the {fundamental} constraint (\ref{eq:linksim}) requiring that each cell vertex should be the orthocentre of the triangle formed by adjacent edge centroids (or, equivalently, of the triangle formed by the three adjacent vertices), { from which we were able to develop a self-consistent dual network (Appendix~\ref{sec:dual})}.  {Our strategy of using polygonal cell boundaries to define the primal network, and using physical constraints to identify an appropriate dual triangulation (specifically via an orthocentric construction), differs from many other studies in the discrete calculus literature in which a simplicial complex is taken to be primal and \textit{a priori} barycentric or circumcentric constructions are used to build a dual network of polygons \cite{desbrun2005, yavari2008, grady2010}}.

The force network and Airy stress function both provide mechanisms for visualising stress.  
Stress can be interpreted as a map between the centroid network (with edges $\mathbf{s}_{ik}$ and vertices $\mathbf{c}_j$) and the force network (with edges $-\boldsymbol{\varepsilon}\mathbf{f}_{ik}$ and vertices $\mathbf{h}_j$).   However this map can be distorted, with the periphery of the force network for example shrinking to zero as the external load $P_{\mathrm{ext}}$ tends to zero.  The isotropic stress fields, $P_{\mathrm{eff},i}$ over cells or $P_{\mathrm{eff},k}$ over tristars (\ref{eq:peffilap}, \ref{eq:peffk}), are defined as discrete Lapalacians 
 \begin{equation}
\mathsf{G} \mathsf{A}^T\mathsf{T}\mathsf{A}, \quad  \mathsf{B}\mathsf{T}^{-1}\mathsf{B}^T \mathsf{H},
 \label{eq:lapl}
 \end{equation}
of the Airy stress function,  where $\mathsf{G}\equiv \mathrm{diag}(1/E_k)$, $\mathsf{H}\equiv \mathrm{diag}(1/A_i)$, $\mathsf{T}\equiv \mathrm{diag}(T_j/t_j)$ serve the role of Hodge star operators \cite{grady2010}.  
Ramola \& Chakraborty \cite{ramola2017} used the spectral properties of a graph Laplacian as a tool to understand force localisation in granular materials.  Likewise, the geometrically-weighted Laplacians (\ref{eq:lapl}), defined on the vertices of the primal and dual networks, are promising candidates for determining the structure of mesoscopic patterns of stress in cellular materials, one of a class of potentially significant mechanical heterogeneities \cite{blanchard2019}.  {However further work is needed to identify the analogue in the present problem (if it exists) of the Beltrami-Michell equation (which leads to the Airy stress function satisfying a biharmonic equation in continuous planar elasticity), which would make $P_{\mathrm{eff}}$ harmonic.}  A further useful visualisation arises from the constraint that the orientation of intracellular stress $\boldsymbol{\sigma}_k$ in the tristar that surrounds vertex $k$ must share its principal axes with the fabric tensor $\mathsf{F}_k$, provided there is sufficient asymmetry for $\mathsf{F}_k$ to be well defined.  An analogous construction in granular materials has been connected to the orientation of force chains \cite{blumenfeld2004}.  Remarkably, the stress-geometry condition does not depend directly on the choice of constitutive model.   

In simulating the vertex model, it is common to either minimize a total mechanical energy $U(\mathbf{r})$ directly by displacing the vertices $\mathbf{r}_k$, or to apply a drag $\eta$ to each vertex, so that an equilibrium is reached by timestepping $N_v$ coupled ODEs for $\mathbf{r}_k(t)$ of the form (\ref{eq:verdyn}).
In both approaches, the cell stress $\boldsymbol{\sigma}_i$ can be evaluated and, happily, it is symmetric (\ref{eq:cellstress}), ensuring local torque balance.  However, this condition is not sufficient to ensure global torque balance, as consideration should also be given to the stress ${\boldsymbol{\sigma}}_k$ around vertices. 
In other words, our study shows that cellular materials described by the vertex model should also be subject to a stress-geometry condition (\ref{eq:fes}) equivalent to that described for granular materials \cite{ball2002, degiuli2014}.  Our study therefore suggests that it is necessary to constrain the optimisation of $U$ (for example via candidate algorithm (\ref{eq:opt})), to ensure that appropriate geometric constraints are satisfied as the system approaches a final  equilibrium state.  A {secondary} construction identifying cell centres { (Appendix~\ref{sec:dual})}, {allows} imposition of (\ref{eq:orthog}).  We will address computational approaches with which to implement the torque balance conditions (\ref{eq:orthog}, \ref{eq:linksim}) elsewhere.  

{
This study is based on two fundamental assumptions: first, that the forces acting on each vertex can be partitioned into contributions from each cell, and that these constitute all the forces in the system; second, that there are no intra- or inter-cellular torques.  From these assumptions, we deduced orthogonality of links and edges, and orthocentricity of vertices with respect to their neighbours.    An alternative strategy was taken by \cite{noll2017,noll2018}, who partitioned forces at vertices (such as (\ref{eq:forver})) into contributions from each edge, deriving a triangulated dual network embedded in $\mathbb{R}^3$.  The relationship between these approaches is discussed briefly in Appendix~\ref{sec:foredge}.  An interesting further consequence of vertex orthocentricity is that all internal angles of polygonal cells should exceed $\pi/2$, implying that a T1 transition (a neighbour exchange) will arise as soon as the internal angle between adjacent cell edges becomes too acute.  This is in contrast to the standard vertex model, when a threshold condition is often needed to trigger such a transition \cite{spencer2017}.}

As Fig.~\ref{fig:ex}(d) illustrates, orthogonality between links (connecting cell centroids) and edges (connecting vertices) is imperfect in real systems.  There are obvious epistemic reasons: there are errors in the measurement and segmentation of cell boundaries; cell walls are not straight; and additional forces acting on some cells (due to division or motility), that are not easily partitioned at vertices, may be missing from the force balance (\ref{eq:foreq}).  Further, while cell centroids can be determined directly from images (using (\ref{eq:centroid2})), these {will typically} deviate from the cell centres that enable conditions such as (\ref{eq:linksim}) to be satisfied.  Careful optimisation strategies will therefore be needed to align self-consistent models that respect torque balance with data such as Fig.~\ref{fig:ex}.  It also remains to be seen to what extent the neglect of global torque balance has influenced predictions of previous computational realisations of the vertex model.  
The discrepancy may be subdominant to many of the other approximations implicit in modelling complex biological cells with simple physical models.  For example, models that impose a Voronoi structure on the monolayer, solving only for the motion of cell centres, gain computational efficiency at the cost of some fidelity \cite{barton2017, yang2017}, at a level that has previously been judged acceptable for the purposes of the studies in question.  Nevertheless it is desirable to ensure physical balances are properly and fully respected, particularly as models grow in sophistication, {and we argue that an orthocentric construction is more appropriate}.  At a more fundamental level, the appearance of a stress-geometry condition also raises intriguing questions about the role of microstructure in homogenized models of biological tissues. 

In summary, by identifying the underlying structure of the stress field {implicit in the vertex model} in terms of an Airy stress function and { by identifying geometric constraints arising from torque balance}, this study supports the development of more robust simulations, facilitates deeper understanding of the mesoscopic structures in disordered cellular monolayers, and provides a secure foundation for future upscaling approximations.
 
 
 
 
 
 \section*{Acknowledgements}
 {SW was supported by a Wellcome Trust/Royal Society Sir Henry Dale Fellowship [098390/Z/12/Z] and EJ by a Wellcome Trust 4-yr PhD studentship [210062/Z/17/Z]. Conversations with Alex Nestor-Bergmann are gratefully acknowledged.}
 
 
\begin{appendix}

\section{Incidence matrices}
\label{sec:bo}

\begin{figure}
\begin{center}
\includegraphics[height=3.5in]{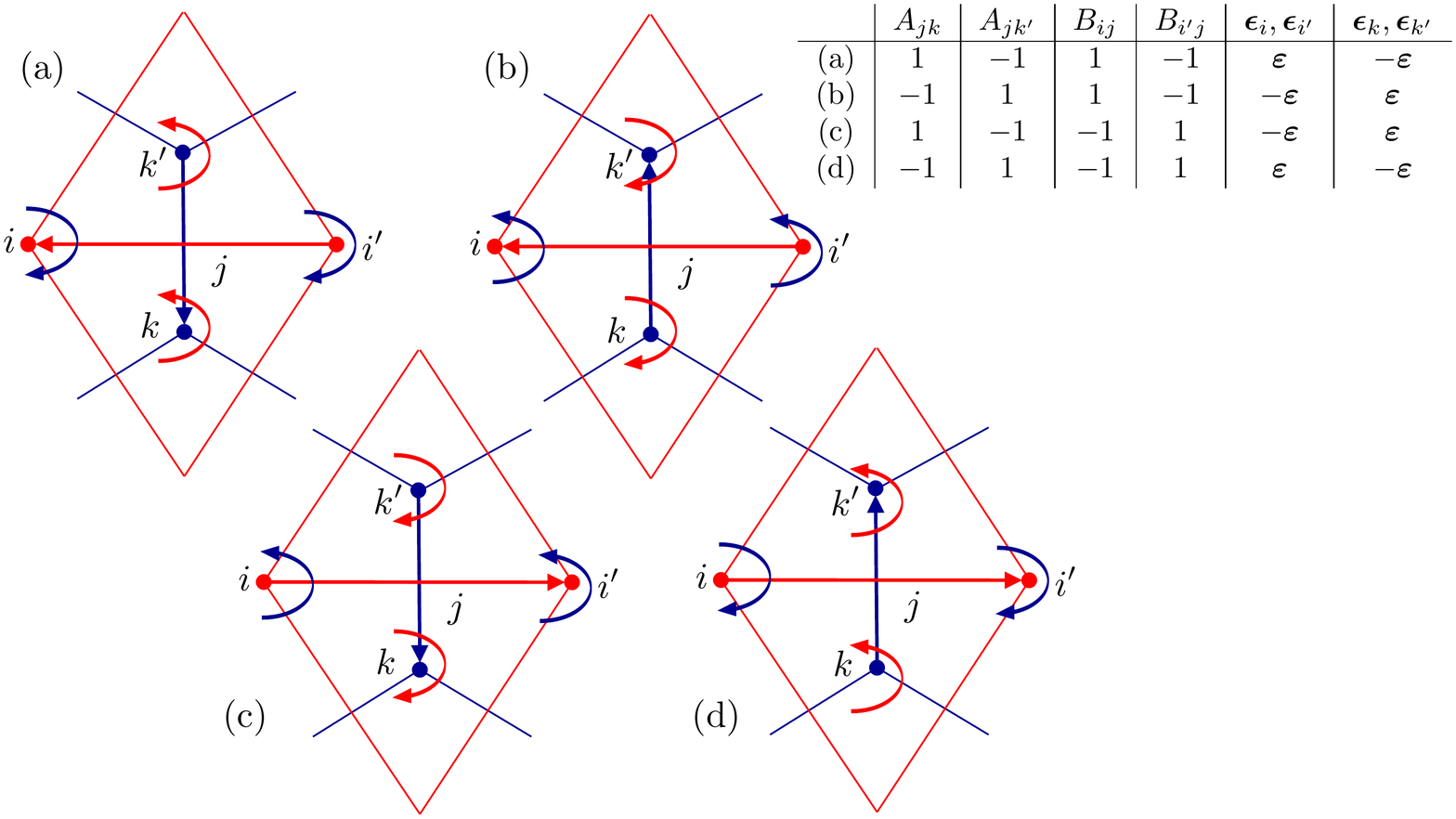}
\end{center}
\caption{ Four different cases having consistent orientations of the link $\mathbf{t}_j$ (blue) between cell vertices $k$ and $k'$, and the link $\mathbf{T}_j$ (red) between cell centres $i$ and $i'$.  The corresponding orientations of cells $i$ and $i'$, and triangles $k$ and $k'$ are also shown.  The tablular inset shows corresponding values of incidence matrices.  Cell and triangle orientations are given in terms of the Levi-Civita symbol $\boldsymbol{\varepsilon}$, representing clockwise $\pi/2$ rotation.  }
\label{fig:pridu}
\end{figure}

Treating the network of cells as primal, then a dual network is the triangulation between adjacent cell centres \cite{grady2010}.  An orientation imposed on one network induces an orientation on the other.  As Fig.~\ref{fig:pridu} 
illustrates, consistency requires cells to share the same orientation, and triangles to share the opposite orientation.  Consider cell edge $\mathbf{t}_j$, connecting vertices $k$ and $k'$, that is dual to the link $\mathbf{T}_j$ between the centres of cells $i$ and $i'$.  All possible values of the entries in $\mathsf{A}$ (indicating the orientation of $\mathbf{t}_j$ with respect to vertices $k$ and $k'$ and of triangles with respect to $\mathbf{T}_j$) and $\mathsf{B}$ (indicating the orientation of $\mathbf{T}_j$ with respect to vertices $i$ and $i'$ and of cells with respect to $\mathbf{t}_j$) are enumerated in the figure.  

$\mathsf{A}^T$ and $\mathsf{B}^T$ have interpetations as boundary operators.  Since all cells have edges that form  closed cycles, $\mathsf{A}^T\mathsf{B}^T=\mathsf{0}$, and hence $\mathsf{B}\mathsf{A}=\mathsf{0}$ \cite{grady2010}.  The rank-nullity theorem gives
$\mathrm{dim}\left(\mathrm{ker} (\mathsf{A}^T)\right)+\mathrm{dim}\left(\mathrm{im} (\mathsf{A}^T)\right)=N_e$ and 
$\mathrm{dim}\left(\mathrm{ker} (\mathsf{B}^T)\right)+\mathrm{dim}\left(\mathrm{im} (\mathsf{B}^T)\right)=N_c$.
$\mathrm{ker} (\mathsf{A}^T)$ identifies sets of edges with no boundary, \hbox{i.e.} edges that form closed cycles.  The $N_c$ cell boundaries form a linearly independent basis for all such cycles, and hence $\mathrm{dim}(\mathrm{ker} (\mathsf{A}^T))=N_c$.  For a localized monolayer, there is no combination of cells that has no boundary, and so $\mathrm{dim}(\mathrm{ker} (\mathsf{B}^T))=0$.  It follows that $\mathsf{B}^T$, and hence $\mathsf{B}$, is full rank ($\mathrm{dim}(\mathrm{im} (\mathsf{B}^T))=N_c$, \hbox{i.e.} the set of all cell boundaries is spanned by $N_c$ independent cycles), whereas $\mathsf{A}$ has rank $\mathrm{dim}(\mathrm{im} (\mathsf{A}^T))=\mathrm{dim}(\mathrm{im} (\mathsf{A}))=N_e-N_c$ (giving the size of the set of independent boundaries of edges).  



\section{Links between edge centroids}
\label{sec:cik}

To establish (\ref{eq:cik}), consider two edges of cell $i$, $j$ and $j'$, meeting at vertex $k$, with $j$ preceding $j'$ when listed clockwise.  There are four possible orientations of the edges, with $(A_{jk}, A_{j'k})=(1,-1)$, $(-1,-1)$, $(1,1)$ and $(-1,1)$.  If $\boldsymbol{\epsilon}_i=\boldsymbol{\varepsilon}$, then the corresponding values of $(B_{ij}, B_{ij'})$ are $(1,1)$, $(-1,1)$, $(1,-1)$ and $(-1,-1)$.  In all cases, the product $(B_{ij} A_{jk}, B_{ij'} A_{j'k})$ is $(1,-1)$.  Since $\boldsymbol{\varepsilon}\boldsymbol{\epsilon}_i=-\mathsf{I}$ in this case, then $\boldsymbol{\varepsilon}\boldsymbol{\epsilon}_i \sum_j B_{ij} \mathbf{c}_j A_{jk}=\mathbf{c}_{j'}-\mathbf{c}_j$.  Altneratively, if $  \boldsymbol{\epsilon}_i=-\boldsymbol{\varepsilon}$, the sign change of $B_{ij}$ cancels with the sign changes of $\boldsymbol{\varepsilon}\boldsymbol{\epsilon}_i$, giving the same result, hence establishing (\ref{eq:cik}).

To establish (\ref{eq:fabric}), consider the three cells $i$, $i'$, $i''$ arranged anticlockwise around vertex $k$, separated by edges $j$, $j'$ and $j''$ as indicated in Fig.~\ref{fig:airy}.  Suppose first that $\boldsymbol{\epsilon}_i=\boldsymbol{\varepsilon}$ and $\boldsymbol{\epsilon}_k=-\boldsymbol{\varepsilon}$.  There are then 8 possible values of $A_{jk}$, namely $(1, 1, 1)$, $(1, -1, 1)$, $(1, 1, -1)$, $(1, -1, -1)$, $(-1, -1, -1)$, $(-1, 1, -1)$, $(-1, -1, 1)$, $(-1, 1, 1)$, reflecting the orientation of edges.  Suppose edge $j$ points inwards.  Then $B_{ij}=1$ and $B_{i'' j}=-1$, so that $B_{ij}A_{jk}$ takes the values $-1$ and $1$ when summed over cells in the anticlockwise direction.  Likewise, suppose edge $j''$ points outwards.  Then $B_{i'j''}=1$ and $B_{i'' j''}=-1$, so that $B_{ij}A_{jk}$ again takes the values $-1$ and $1$ when summed over cells in the anticlockwise direction.  Thus $\boldsymbol{\varepsilon}\boldsymbol{\varepsilon}_k \sum_{i,j} B_{ij} A_{jk}$ produces the signature $(1,-1)$, $(1,-1)$,  $(1,-1)$ when taken over the three cells surrounding vertex $k$, looping around the six edges of the tristar.   Assuming instead that $\boldsymbol{\epsilon}_k=\boldsymbol{\varepsilon}$ leads to reversals in the sign of $\boldsymbol{\varepsilon}\boldsymbol{\varepsilon}_k$ and of $B_{ij}$, which cancel, leading to the same pattern, hence establishing (\ref{eq:fabric}).

\section{Kites}
\label{sec:kites}

\begin{figure}
\begin{center}
\includegraphics[width=5.5in]{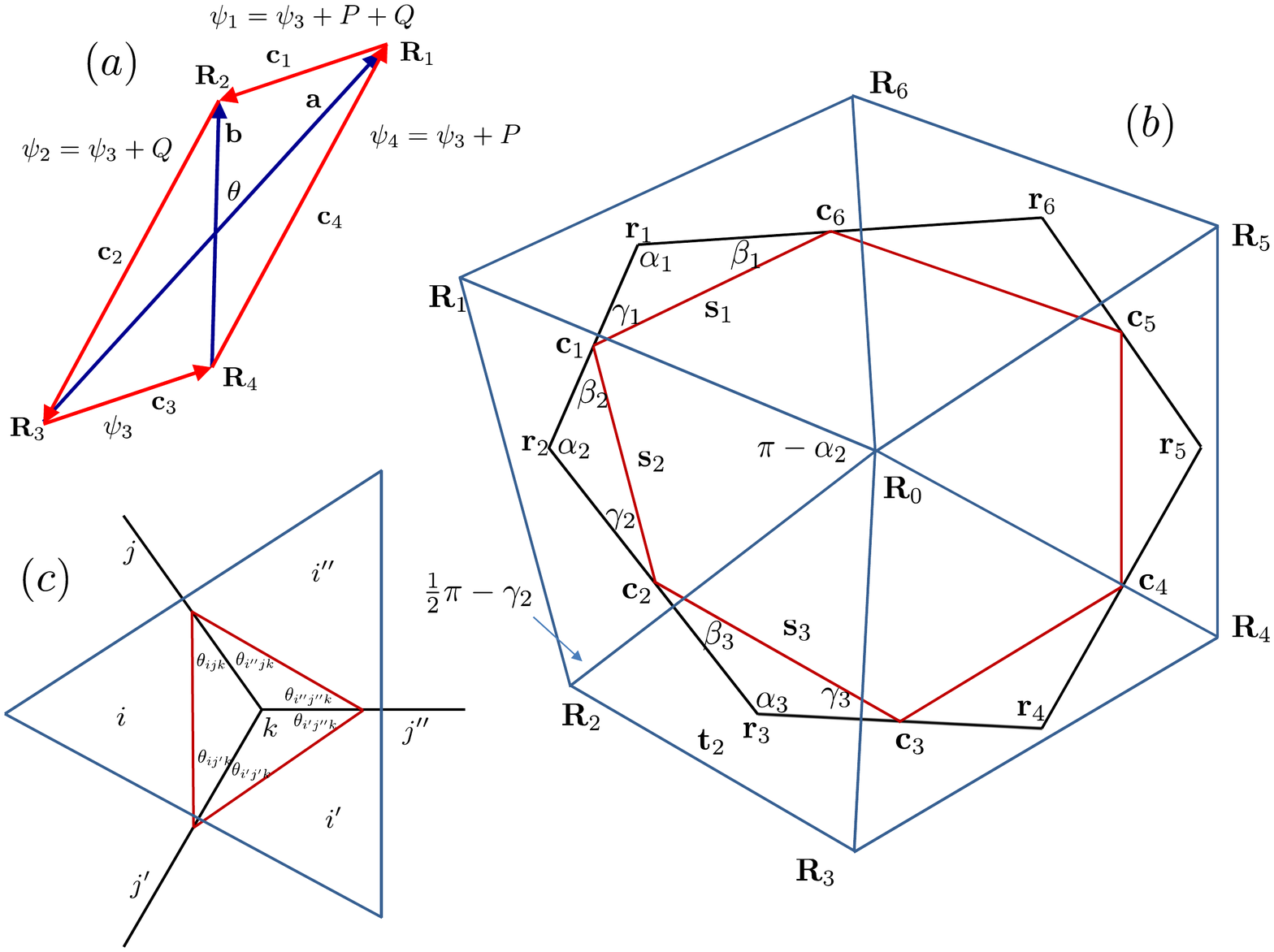}
\end{center}
\caption{(a) For a vector $\mathbf{g}$ to be written as a discrete curl along the path given by $\mathbf{c}_1$, $\mathbf{c}_2$, $\mathbf{c}_3$, $\mathbf{c}_4$, the jumps in scalar potential $P$ and and $Q$ are given by the coefficients of $\mathbf{b}$ and $-\mathbf{a}$ in (\ref{eq:b4}). {(b) Construction of a triangulation dual to a cell having prescribed vertex locations, defined up to the cell centre location $\mathbf{R}_0$ and the length of one link.  (c)  Around vertex $k$, the ratios of adjacent sides of the orthogonal triangulation are given by ratios of cosines of the six interior angles of the triangle connecting edge centroids. }
 }
\label{fig:kite}
\end{figure}

Consider the quadrilateral defined by non-parallel vectors $\mathbf{a}$ and $\mathbf{b}$ (Fig.~\ref{fig:kite}a).  Imagine they intersect at the origin, and that the vertices $\mathbf{R}_i$, ($i=1,2,3,4$, ordered anticlockwise) satisfy $\mathbf{a}=\mathbf{R}_1-\mathbf{R}_3$ and $\mathbf{b}=\mathbf{R}_2-\mathbf{R}_4$.  Let the edge vectors forming its perimeter be $\mathbf{c}_i=\mathbf{R}_{i+1}-\mathbf{R}_i$ (writing $\mathbf{R}_5\equiv\mathbf{R}_1$), looping anticlockwise around the quadrilateral.  Then the symmetric component of $\mathbf{a}\otimes\mathbf{b}$ satisfies
\begin{equation}
\tfrac{1}{2}(\mathbf{a}\mathbf{b}^T+\mathbf{b}\mathbf{a}^T)=-\tfrac{1}{2}(\mathbf{c}_1\mathbf{c}_1^T-\mathbf{c}_2\mathbf{c}_2^T+\mathbf{c}_3\mathbf{c}_3^T-\mathbf{c}_4\mathbf{c}_4^T).
\label{eq:quad}
\end{equation}
This can be demonstrated by substitution, writing each side of (\ref{eq:quad}) in terms of $\mathbf{R}_i$.  The antisymmetric component satisfies
$\tfrac{1}{2}(\mathbf{a}\mathbf{b}^T-\mathbf{b}\mathbf{a}^T)=A \boldsymbol{\varepsilon},$
where $A=\tfrac{1}{2} \vert\mathbf{a}\vert \vert\mathbf{b}\vert \sin\theta$ is the area of the quadrilateral and $\theta$ is the angle between $\mathbf{a}$ and $\mathbf{b}$.  Thus, following \cite{ball2002}, we can write
\begin{equation}
\mathbf{a}\otimes \mathbf{b}= A\boldsymbol{\varepsilon} -\tfrac{1}{2}(\mathbf{c}_1\otimes \mathbf{c}_1-\mathbf{c}_2\otimes \mathbf{c}_2+\mathbf{c}_3\otimes \mathbf{c}_3-\mathbf{c}_4\otimes \mathbf{c}_4).
\end{equation}

Now suppose that $\mathbf{g}=\psi_1\mathbf{c}_1 + \psi_2 \mathbf{c}_2 + \psi_3 \mathbf{c}_3 + \psi_4 \mathbf{c}_4$ (a discrete curl of a potential) and that $\psi_1-\psi_2=\psi_4-\psi_3=P$ and $\psi_2-\psi_3=\psi_1-\psi_4=Q$ (Fig.~\ref{fig:kite}a).   Then eliminating $\psi_2$ gives 
$\mathbf{g}=\psi_1\mathbf{c}_1 + (\psi_1-P) \mathbf{c}_2 + \psi_3 \mathbf{c}_3 + (\psi_1+P) \mathbf{c}_4$ with $\psi_1-\psi_3=P+Q$.  Eliminating $\psi_3$ then gives $\mathbf{g}=P\mathbf{b}-Q\mathbf{a}$.  Projecting this onto $\mathbf{a}$ and $\mathbf{b}$ and rearranging (assuming $\mathbf{a}$ and $\mathbf{b}$ are not parallel) gives
\begin{equation}
\mathbf{g}=\left[\frac{(\mathbf{g}\cdot\mathbf{b})(\mathbf{a}\cdot\mathbf{a})-(\mathbf{g}\cdot\mathbf{a})(\mathbf{a}\cdot\mathbf{b})}
 {(\mathbf{a}\cdot\mathbf{a})(\mathbf{b}\cdot \mathbf{b})-(\mathbf{a}\cdot\mathbf{b})^2}\right] \mathbf{b}+
 \left[ \frac{(\mathbf{g}\cdot\mathbf{a})(\mathbf{b}\cdot\mathbf{b})-(\mathbf{g}\cdot\mathbf{b})(\mathbf{a}\cdot\mathbf{b})}
 {(\mathbf{a}\cdot\mathbf{a})(\mathbf{b}\cdot \mathbf{b})-(\mathbf{a}\cdot\mathbf{b})^2} 
\right] \mathbf{a}.
\label{eq:b4}
\end{equation}
For the jumps in potential $P$ and $Q$ to depend independently on $\mathbf{g}\cdot\mathbf{a}$ or $\mathbf{g}\cdot\mathbf{b}$ then requires $\mathbf{a}\cdot\mathbf{b}=0$, in which case (\ref{eq:b4}) gives the familiar orthogonal projection $P=(\mathbf{g}\cdot\mathbf{b})\mathbf{b}/b^2$ and $Q=-(\mathbf{g}\cdot\mathbf{a})\mathbf{a}/a^2$. 

Finally, for a vector $\mathbf{h}$, we seek the potential such that the jump $P$ in the $\mathbf{a}$ direction is $\mathbf{h}\cdot\mathbf{a}$ and that the jump $Q$ in the $\mathbf{b}$ direction is $\mathbf{h}\cdot\mathbf{b}$.  The necessary construction is a discrete curl around the periphery of the rectangle bounded by $\mathbf{a}$ and $\mathbf{b}$, namely
\begin{subequations}
\begin{equation}
\mathbf{h}=\frac{\boldsymbol{\varepsilon}}{ab} (\psi_1\mathbf{c}_1 + \psi_2 \mathbf{c}_2 + \psi_3 \mathbf{c}_3 + \psi_4 \mathbf{c}_4)=\frac{\boldsymbol{\varepsilon}}{ab} (P\mathbf{b} - Q \mathbf{a})
=\frac{(\mathbf{h}\cdot \mathbf{a})\mathbf{a}}{a^2} + \frac{(\mathbf{h}\cdot\mathbf{b}) \mathbf{b}}{b^2}
\end{equation}
when $\boldsymbol{\varepsilon} \mathbf{b}=(b/a)\mathbf{a}$, $\boldsymbol{\varepsilon} \mathbf{a}=-(a/b)\mathbf{b}$.  The same construction applies when $\boldsymbol{\varepsilon} \mathbf{b}=-(b/a)\mathbf{a}$, $\boldsymbol{\varepsilon} \mathbf{a}=(a/b)\mathbf{b}$ and 
\begin{equation}
\mathbf{h}=\frac{\boldsymbol{\varepsilon}}{ab} \left[(\mathbf{h}\cdot\mathbf{b}) \mathbf{a} - (\mathbf{h}\cdot\mathbf{a}) \mathbf{b} \right].
\end{equation}
\label{eq:curot}
\end{subequations}

\section{Force network area}
\label{sec:areamap}

The mapping $\mathsf{M}_k$ is linear and uniform over the triangle surrounding vertex $k$.  Thus $\mathbf{f}_{ik}=\boldsymbol{\varepsilon}\mathsf{M}_k \mathbf{s}_{ik}=\boldsymbol{\sigma}_k\boldsymbol{\varepsilon} \mathbf{s}_{ik}$.  This map takes  $\mathbf{d}_{ik}=\tfrac{1}{2}\sum_j \overline{A}_{jk} \overline{B}_{ij}\mathbf{c}_j$, the midpoint of $\mathbf{s}_{ik}$, to $\mathbf{e}_{ik}=\tfrac{1}{2}\sum_j \overline{A}_{jk} \overline{B}_{ij}\mathbf{h}_j$, the midpoint of $\mathbf{f}_{ik}$.  The oriented area of the triangle around vertex $k$ is
$W_k\boldsymbol{\epsilon}=-{\textstyle{\sum_i} } {C}_{ik} \mathbf{s}_{ik}\otimes \mathbf{d}_{ik}$.
Then the area $V_k$ of the corresponding triangle in the force network satisfies
\begin{align}
V_k \mathsf{I}=& {\textstyle{\sum_i} } {C}_{ik}( \mathbf{f}_{ik}\otimes \mathbf{e}_{ik})=
 { \textstyle{\sum_i} } {C}_{ik} (\boldsymbol{\sigma}_k\boldsymbol{\varepsilon}\mathbf{s}_{ik})\otimes (-\boldsymbol{\varepsilon}\boldsymbol{\sigma}_k \boldsymbol{\varepsilon}\mathbf{d}_{ik}) \nonumber \\
& =\boldsymbol{\sigma}_k\boldsymbol{\varepsilon} \left[
{ \textstyle{\sum_i} }{C}_{ik} \mathbf{s}_{ik} \otimes\mathbf{d}_{ik}\right]
(-\boldsymbol{\varepsilon}\boldsymbol{\sigma}_k \boldsymbol{\varepsilon})^T 
=- W_k \boldsymbol{\sigma}_k\boldsymbol{\varepsilon} \boldsymbol{\sigma}_k \boldsymbol{\varepsilon} = W_k \,\mathrm{det}(\boldsymbol{\sigma}_k) \mathsf{I},
\end{align}
using the identity $\boldsymbol{\varepsilon}\mathsf{A}\boldsymbol{\varepsilon} \mathsf{A}^T=- \mathrm{det}(\mathsf{A})\mathsf{I} $.  Thus $V_k=\mathrm{det}(\boldsymbol{\sigma}_k) W_k$.


\section{Relating Airy stress function to cell shape}
\label{sec:air}

Let $\mathbf{1}_i=\{1\}$, $\mathbf{v}_i=\{\overline{B}_{ij} v_{ij}\}$, $\mathbf{t}_i=\{\overline{B}_{ij} t_j/L_i\}$, $\mathbf{c}_i=\{\overline{B}_{ij} \cos 2\alpha_j \}$, $\mathbf{s}_i=\{ \overline{B}_{ij} \sin 2\alpha_j\}$ where the vectors $\{ \cdot \}$ gather $Z_i$ non-zero elements into a vector over edges of cell $i$.  We use the dot product to represent scalar products between these vectors, so that $\mathbf{t}_i\cdot\mathbf{1}_i=1$ and $\mathbf{1}_i\cdot\mathbf{1}_i=Z_i$.  Then the deviatoric components of stress tensors in (\ref{eq:gi}), (\ref{eq:cellstress}) match if
\begin{equation}
\mathbf{v}_i\cdot \mathbf{1}_i=0, \quad \mathbf{v}_i\cdot\mathbf{s}_i=\pm \mathcal{T}_i L_i \mathbf{t}_i\cdot\mathbf{c}_i, \quad \mathbf{v}_i\cdot\mathbf{c}_i=\mp \mathcal{T}_i L_i \mathbf{t}_i\cdot \mathbf{s}_i \quad\mathrm{for}\quad \boldsymbol{\epsilon}_i=\mp\boldsymbol{\varepsilon}.
\label{eq:vcond}
\end{equation}
Writing $\mathbf{v}_i=\pm \mathcal{T}_iL_i (\alpha_i \mathbf{1}_i +\beta_i\mathbf{c}_i +\gamma_i\mathbf{s}_i)$, the three scalar conditions (\ref{eq:vcond}) determine $\alpha_i$, $\beta_i$ and $\gamma_i$ as 
\begin{subequations}
\begin{align}
Z_i \alpha_i  &= -\beta_i (\mathbf{c}_i\cdot\mathbf{1}_i)-\gamma_i (\mathbf{s}_i\cdot \mathbf{1}_i), \\
\beta_i \Delta_i &= (\mathbf{t}_i\cdot\mathbf{c}_i) Z_i [Z_i  (\mathbf{s}_i\cdot\mathbf{c}_i) - (\mathbf{s}_i\cdot\mathbf{1}_i) (\mathbf{c}_i\cdot\mathbf{1}_i)] + (\mathbf{t}_i\cdot\mathbf{s}_i) Z_i [Z_i (\mathbf{s}_i\cdot\mathbf{s}_i) - (\mathbf{s}_i\cdot\mathbf{1}_i)^2], \\
\gamma_i \Delta_i &= - (\mathbf{t}_i\cdot\mathbf{c}_i) Z_i[Z_i (\mathbf{c}_i\cdot\mathbf{c}_i) -(\mathbf{c}_i\cdot\mathbf{1}_i)^2 ] - (\mathbf{t}_i\cdot\mathbf{s}_i)Z_i  [Z_i (\mathbf{c}_i\cdot\mathbf{s}_i) -(\mathbf{c}_i\cdot\mathbf{1}_i) (\mathbf{s}_i\cdot\mathbf{1}_i)], \\
\Delta_i &=  Z_i^2 [(\mathbf{c}_i\cdot\mathbf{s}_i)^2 - (\mathbf{c}_i\cdot\mathbf{c}_i) (\mathbf{s}_i\cdot\mathbf{s}_i) ] + Z_i [(\mathbf{c}_i\cdot\mathbf{c}_i) (\mathbf{s}_i\cdot\mathbf{1}_i)^2  + (\mathbf{s}_i\cdot\mathbf{s}_i) (\mathbf{c}_i\cdot\mathbf{1}_i)^2 \nonumber \\ & \qquad \qquad \qquad \qquad \qquad \qquad \qquad \qquad \qquad \qquad \qquad 
- 2 (\mathbf{s}_i\cdot\mathbf{c}_i) (\mathbf{s}_i\cdot\mathbf{1}_i) (\mathbf{c}_i\cdot\mathbf{1}_i)].
\end{align}
\label{eq:airyjump}
\end{subequations}
Given the jumps $\mathbf{v}_i$, the values of the Airy stress in cell $i$, $\boldsymbol{\psi}_i$, can be expressed in terms of a mean value $\overline{\psi}_i$ as 
\begin{equation}
\left(\begin{matrix}
\psi_{i1} \\ \psi_{i2} \\ \psi_{i3} \\ \dots \\ \psi_{i,Z_i-1} \\ \psi_{i,Z_i}
\end{matrix}\right)
=\overline{\psi} 
\left(\begin{matrix}
1 \\ 1 \\ 1 \\ \dots \\ 1 \\1
\end{matrix}\right)
+\frac{1}{Z_i} 
\left(
\begin{matrix}
1-Z_i & 2-Z_i & 3-Z_i & \dots & -2 & -1 \\
1 & 2-Z_i & 3-Z_i & \dots & -2 & -1 \\
1 &  2 & 3-Z_i & \dots & -2 & -1 \\
\dots \\
1 & 2 & 3 & \dots & Z_i-2 & -1 \\
1 & 2 & 3 & \dots & Z_i-2 & Z_i-1 
\end{matrix}
\right)
\left(
\begin{matrix}
v_{i1} \\ v_{i2} \\ v_{i3} \\ \dots \\ v_{i,Z_i-2} \\ v_{i,Z_i-1}
\end{matrix}
\right).
\end{equation}
Thus the variations in Airy stress function in cell $i$ are given by $\mathcal{T}_i L_i$ times a dimensionless function of cell shape.  Noting that $\mathrm{det}(\mathsf{D}_i)=-\tfrac{1}{4} [(\mathbf{v}_i\cdot\mathbf{c}_i)(\mathbf{v}_i\cdot\mathbf{c}_i)+(\mathbf{v}_i\cdot\mathbf{s}_i)(\mathbf{v}_i\cdot\mathbf{s}_i)]=\mathrm{det}(\mathsf{Q}_i)$, the magnitude of the shear in cell $i$ is $\zeta = \tfrac{1}{2}\sqrt{(\mathbf{v}_i\cdot\mathbf{c}_i)^2+(\mathbf{v}_i\cdot\mathbf{s}_i)^2}
= \tfrac{1}{2}\mathcal{T}_i L_i \sqrt{(\mathbf{t}_i\cdot\mathbf{c}_i)^2+(\mathbf{t}_i\cdot\mathbf{s}_i)^2}$.  

The relation (\ref{eq:airyjump})  between cell shape $(\mathbf{c}_i, \mathbf{s}_i)$ and intracellular difference in Airy stress function $(\mathbf{v}_i)$ places further constraints on the shapes of neighbouring cells.  Consider for example two cells (say $i$ and $i'$) with a common edge, and the four kites having this edge as part of their boundary.  The intracellular jump in Airy stress function between two neighbouring kites (\ref{eq:airyjump}) is the same in both cells.  Thus, knowing the shape of each cell specifies the ratio $(\mathcal{T}_i L_i)/(\mathcal{T}_{i'} L_{i'})$.   Extending this argument to cells $i$, $i'$ and $i''$ sharing a common vertex (as in Fig.~\ref{fig:airy}), the ratios $(\mathcal{T}_{i'} L_{i'})/(\mathcal{T}_{i''} L_{i''})$ and $(\mathcal{T}_{i''} L_{i''})/(\mathcal{T}_{i} L_{i})$ are also specified by cell shapes.  The product of the three ratios gives a nonlinear shape constraint on the three cells, arising effectively from imposing a torque balance on the tristar bounded by the three cell centres.


{
\section{Construction of the dual network}\label{sec:dual}

We show below how an orthocentric primal network admits a dual orthogonal triangulation, supporting (\ref{eq:opt}) as the basis of a computational scheme.  


Consider a cell with vertices $\mathbf{r}_1, \dots, \mathbf{r}_Z$, edges $\mathbf{t}_i=\mathbf{r}_{i+1}-\mathbf{r}_i$ (taking $\mathbf{t}_{Z+1}\equiv \mathbf{t}_1$ etc.), edge centroids $\mathbf{c}_i=\tfrac{1}{2} (\mathbf{r}_i+\mathbf{r}_{i+1})$ and links between them $\mathbf{s}_i=\mathbf{c}_{i}-\mathbf{c}_{i-1}$ (Fig.~\ref{fig:kite}b).  Choose a cell centre $\mathbf{R}_0$ and radiate lines from $\mathbf{R}_0$ that are orthogonal to each edge.  Choose a point $\mathbf{R}_1$ on the line crossing $\mathbf{t}_1$.  Then construct continuous straight line segments that are parallel to $\mathbf{s}_2, \mathbf{s}_3, \dots, \mathbf{s}_Z$, intersecting the radiating lines at $\mathbf{R}_2, \dots, \mathbf{R}_Z$ respectively.  We wish to establish if the remaining link from $\mathbf{R}_Z$ to $\mathbf{R}_1$ is parallel to $\mathbf{s}_1$, forming a closed loop of links that are orthogonal to edges.   Now given $\mathbf{R}_i$, for $\mathbf{R}_{i+1}-\mathbf{R}_0$ to be orthogonal to $\mathbf{t}_{i+1}$ we require
\begin{equation}
\mathbf{R}_{i+1}=\mathbf{R}_i-\frac{(\mathbf{R}_i-\mathbf{R}_0)\cdot \mathbf{t}_{i+1}}{\mathbf{s}_{i+1}\cdot \mathbf{t}_{i+1}} \mathbf{s}_{i+1}\equiv \mathbf{R}_i+\vert \mathbf{R}_i-\mathbf{R}_0\vert \frac{\sin\alpha_{i+1}}{\cos\gamma_{i+1}} \hat{\mathbf{s}}_{i+1}
\end{equation}
where $\alpha_i$ is the internal angle between edges at vertex $i$ and $\gamma_i$ is the acute angle between $\mathbf{s}_i$ and $\mathbf{t}_i$.  Let $\beta_i=\pi-\gamma_i-\alpha_i$ be the acute angle between $\mathbf{s}_i$ and $\mathbf{t}_{i-1}$.  Note that $\sum_i (\gamma_i+\beta_i)=2\pi$ and all the angles are acute, so that $\cos\gamma_i$ and $\cos\beta_i$ are all positive.  
Noting that $\vert \mathbf{R}_{i+1}-\mathbf{R}_i\vert \cos\gamma_{i+1} = \vert \mathbf{R}_i- \mathbf{R}_0 \vert \sin\alpha_{i+1}$ (the altitude of triangle $\mathbf{R}_0$, $\mathbf{R}_i$, $\mathbf{R}_{i+1}$ through vertex $\mathbf{R}_i$), we have 
\begin{align}
\vert \mathbf{R}_{i+1}-\mathbf{R}_0\vert&= \vert \mathbf{R}_{i+1}-\mathbf{R}_i\vert \cos (\tfrac{1}{2} \pi -\gamma_{i+1})+ \vert \mathbf{R}_i-\mathbf{R}_0 \vert \cos(\pi-\alpha_{i+1}) \nonumber \\
&=\vert \mathbf{R}_i-\mathbf{R}_0 \vert\left[ (\sin\gamma_{i+1} \sin\alpha_{i+1}/\cos\gamma_{i+1})-\cos\alpha_{i+1}\right] \nonumber \\
&=\vert \mathbf{R}_i-\mathbf{R}_0 \vert \cos\beta_{i+1}/\cos\gamma_{i+1}.
\end{align}
Progressing round the cell, the condition 
\begin{equation}
\frac{\cos\beta_1}{\cos\gamma_1} \frac{\cos\beta_2}{\cos\gamma_2}\dots \frac{\cos\beta_Z}{\cos\gamma_Z}=1
\label{eq:a5}
\end{equation}
is therefore required to ensure that the links between $\mathbf{R}_1,\dots,\mathbf{R}_Z$ form a closed loop, with $\mathbf{R}_0$ and $\vert\mathbf{R}_1-\mathbf{R}_0\vert$ as degrees of freedom.

An equivalent construction is illustrated in Fig.~\ref{fig:kite}(c), for which 
\begin{equation}
\frac{T_{j'}}{T_j}=\frac{\cos \theta_{ijk}}{\cos \theta_{ij'k}},\quad
\frac{T_{j''}}{T_{j'}}=\frac{\cos \theta_{i'j'k}}{\cos \theta_{i'j''k}},\quad
\frac{T_{j}}{T_{j''}}=\frac{\cos \theta_{i''j''k}}{\cos\theta_{i''jk}}
\end{equation}
where $\mathbf{T}_j=\vert \mathbf{R}_i-\mathbf{R}_{i''}\vert$, $\mathbf{T}_{j'}=\vert \mathbf{R}_{i'}-\mathbf{R}_{i}\vert$, $\mathbf{T}_{j''}=\vert \mathbf{R}_{i''}-\mathbf{R}_{i'}\vert$.  This demands, for self-consistency,
\begin{equation}
\frac{\cos \theta_{ijk}}{\cos \theta_{ij' k}}\frac{ \cos \theta_{i' j' k}}{\cos \theta_{i'j''k} }\frac{\cos \theta_{i''j''k}}{\cos \theta_{i''jk} }=1.
\end{equation}
which is ensured by symmetry.  This construction allows an orthogonal triangulation to be constructed over an orthocentric primal network, up to translation and uniform scaling.   

To see how these degrees of freedom are accommodated, note first that a mapping $\mathbf{R}_i\mapsto \mathbf{R}_i+\breve{\mathbf{R}}$ for all $i$ implies $\mathbf{T}_j\mapsto \mathbf{T}_j$, $U_k\mapsto U_k$ and therefore $E_k\mapsto E_k$ (via (\ref{eq:tristararea})), implying further that $\boldsymbol{\sigma}_i\mapsto \boldsymbol{\sigma}_i$ and $\boldsymbol{\sigma}_k\mapsto \boldsymbol{\sigma}_k$ and $\psi_{ik}\mapsto \psi_{ik}$.  Translation therefore has no impact on the representation of stress.  Uniform scaling of the dual network is accommodated by jumps in the Airy stress function across cell boundaries.  To see this, recall that $\psi_{ik}$ jumps by $\mathbf{h}_j\cdot \mathbf{t}_j$ between neighbouring kites in a cell that share edge $j$ (this is $\sum_k A_{jk}\theta_{jk}$ in (\ref{eq:thetaphi}a)), and by $\mathbf{h}_j\cdot \mathbf{T}_j$ between adjacent kites in different cells that share edge $j$ (\hbox{i.e.} $\sum_i B_{ij} \phi_{ij}$ in (\ref{eq:thetaphi}b)).  Under the mapping $\mathbf{T}_j\mapsto \mu \mathbf{T}_j$, $U_k\mapsto \mu^2 U_k$, $E_k\mapsto \mu E_k$,  $\sum_k A_{jk}\theta_{jk}\mapsto \sum_k A_{jk}\theta_{jk}$, $\sum_i B_{ij} \phi_{ij}\mapsto \mu \sum_i B_{ij} \phi_{ij}$, it follows that $\mathbf{h}_j\mapsto \mathbf{h}_j$, $\boldsymbol{\sigma}_i\mapsto \boldsymbol{\sigma}_i$ and $\boldsymbol{\sigma}_k\mapsto \boldsymbol{\sigma}_k$.

The orthocentric construction is illustrated in Fig.~\ref{fig:selfsim}(b).  For each cell (black), the condition (\ref{eq:a5}) is satisfied by the linear displacement of a single vertex, allowing cells to be added incrementally.  The dual network of cell centres (blue) was given an arbitrary translation and scaling.

\section{Forces at vertices partitioned by edges}
\label{sec:foredge}

Replacing the sum over $j$ in (\ref{eq:forver}) with a sum over $i$ yields forces $\mathbf{F}_{jk}$ associated with each $j$ edge at vertex $k$.   We indicate here how the force network associated with $\mathbf{F}_{jk}$ can be represented as a triangulated surface embedded in $\mathbb{R}^3$ (consistent with \cite{noll2017, noll2018}) that, when projected onto the $(x,y)$ plane is equivalent to the dual network connecting cell centres (up to translation and scaling), and for which vertices have $z$ coordinates given by cell pressures $\mathcal{P}_i$.  To see this, note that $\mathbf{F}_{jk}$ comprises a force associated with the tension of the two neighbouring cells acting along $\hat{\mathbf{t}}_j$ (of magnitude $\mathcal{T}_i+\mathcal{T}_{i'}$, say), and the pressure difference between the two cells acting normally (of magnitude $\vert \mathcal{P}_i-\mathcal{P}_{i'} \vert t_j/2$).  Rotating the latter vector by $\pi/2$ out of the plane about axis $\hat{\mathbf{t}}_j$, and the former vector in plane by $\boldsymbol{\varepsilon}$, a closed triangle of forces associated with $\cup_{j} \overline{A}_{jk}\mathbf{F}_{jk}$ can be formed lying in a plane that, in general, is not parallel with the physical domain.   However its horizontal projection has edges with lengths $\cup_{j} \overline{A}_{jk} \sum_i  \overline{B}_{ij} \mathcal{T}_{i}$. Consider now the vertex $k'$ at the other end of edge $j$.  It is subject to the same tension and pressure forces, but with opposite orientations, yielding a closed triangle lying in a different plane.  The adjacent edges of the two force triangles stitch together exactly, having the same horizontal projection $\sum_i  \overline{B}_{ij} \mathcal{T}_{i}$ and the same vertical drop $\vert \sum_i B_{ij}  \mathcal{P}_i  t_j/2 \vert$.  It follows that the edges of the dual network connecting cell centres give a direct representation (up to scaling) of the magnitudes and orientations of the composite tension in each cell edge.   
}

\end{appendix}


\bibliographystyle{RS}
\bibliography{abbreviated}

\begin{thebibliography}{99}

\bibitem{ambrosi2019}
Ambrosi D, Ben~Amar M, Cyron CJ, DeSimone A, Goriely A, Humphrey JD, Kuhl E.
  2019  Growth and remodelling of living tissues: perspectives, challenges and
  opportunities. {\em J. Roy. Soc. Interface} \textbf{16}, 20190233.

\bibitem{baskin2013}
Baskin TI, Jensen OE. 2013  On the role of stress anisotropy in the growth of
  stems. {\em J. Exp. Bot.} \textbf{64}, 4697--4707.

\bibitem{alt2017}
Alt S, Ganguly P, Salbreux G. 2017  Vertex models: from cell mechanics to
  tissue morphogenesis. {\em Phil. Trans. R. Soc. B} \textbf{372}, 20150520.

\bibitem{brodland2014}
Brodland GW, Veldhuis JH, Kim S, Perrone M, Mashburn D, Hutson MS. 2014
  CellFIT: a cellular force-inference toolkit using curvilinear cell
  boundaries. {\em PloS One} \textbf{9}, e99116.

\bibitem{farhadifar2007}
Farhadifar R, R{\"o}per JC, Aigouy B, Eaton S, J{\"u}licher F. 2007  The
  influence of cell mechanics, cell-cell interactions, and proliferation on
  epithelial packing. {\em Curr. Biol.} \textbf{17}, 2095--2104.

\bibitem{fletcher2014}
Fletcher AG, Osterfield M, Baker RE, Shvartsman SY. 2014  Vertex models of
  epithelial morphogenesis. {\em Biophys. J.} \textbf{106}, 2291--2304.

\bibitem{fozard2016}
Fozard JA, Bennett MJ, King JR, Jensen OE. 2016  Hybrid vertex-midline
  modelling of elongated plant organs. {\em Interface Focus} \textbf{6},
  20160043.

\bibitem{guirao2015}
Guirao B, Rigaud SU, Bosveld F, Bailles A, Lopez-Gay J, Ishihara S, Sugimura K,
  Graner F, Bella{\"\i}che Y. 2015  Unified quantitative characterization of
  epithelial tissue development. {\em Elife} \textbf{4}, e08519.

\bibitem{ishihara2017}
Ishihara S, Marcq P, Sugimura K. 2017  From cells to tissue: A continuum model
  of epithelial mechanics. {\em Phys. Rev. E} \textbf{96}, 022418.

\bibitem{nagai2001}
Nagai T, Honda H. 2001  A dynamic cell model for the formation of epithelial
  tissues. {\em Phil. Mag. B} \textbf{81}, 699--719.

\bibitem{staple2010}
Staple DB, Farhadifar R, Roeper JC, Aigouy B, Eaton S, J{\"u}licher F. 2010
  {Mechanics and remodelling of cell packings in epithelia}. {\em Eur. Phys. J.
  E} \textbf{33}, 117--127.

\bibitem{weliky1990}
Weliky M, Oster G. 1990  The mechanical basis of cell rearrangement. {I}.
  {E}pithelial morphogenesis during {F}undulus epiboly. {\em Development}
  \textbf{109}, 373--386.

\bibitem{merkel2019}
Merkel M, Baumgarten K, Tighe BP, Manning ML. 2019  A minimal-length approach
  unifies rigidity in underconstrained materials. {\em Proc. Nat. Acad. Sci.}
  \textbf{116}, 6560--6568.

\bibitem{murisic2015}
Murisic N, Hakim V, Kevrekidis I, Shvartsman S, Audoly B. 2015  From Discrete
  to Continuum Models of Three-Dimensional Deformations in Epithelial Sheets.
  {\em Biophys. J.} \textbf{109}, 154 -- 163.

\bibitem{ANB2018b}
Nestor-Bergmann A, Johns E, Woolner S, Jensen OE. 2018  Mechanical
  characterization of disordered and anisotropic cellular monolayers. {\em
  Phys. Rev. E} \textbf{97}, 052409.

\bibitem{bi2015}
Bi D, Lopez J, Schwarz J, Manning ML. 2015  A density-independent rigidity
  transition in biological tissues. {\em Nature Phys.} \textbf{11}, 1074.

\bibitem{bi2016}
Bi D, Yang X, Marchetti MC, Manning ML. 2016  Motility-driven glass and jamming
  transitions in biological tissues. {\em Phys. Rev. X} \textbf{6}, 021011.

\bibitem{boromand2018}
Boromand A, Signoriello A, Ye F, O'Hern CS, Shattuck MD. 2018  Jamming of
  deformable polygons. {\em Phys. Rev. Lett.} \textbf{121}, 248003.

\bibitem{yang2017}
Yang X, Bi D, Czajkowski M, Merkel M, Manning ML, Marchetti MC. 2017
  Correlating Cell Shape and Cellular Stress in Motile Confluent Tissues. {\em
  Proc. Nat. Acad. Sci.} \textbf{114}, 12663--12668.

\bibitem{gao2016}
Gao GJJ, Holcomb MC, Thomas JH, Blawzdziewicz J. 2016  Embryo as an active
  granular fluid: stress-coordinated cellular constriction chains. {\em J.
  Phys.: Condens. Matter} \textbf{28}, 414021.

\bibitem{ANB2018a}
Nestor-Bergmann A, Goddard G, Woolner S, Jensen OE. 2018  Relating cell shape
  and mechanical stress in a spatially disordered epithelium using a
  vertex-based model. {\em Math. Med. Biol.} \textbf{35}, 1--27.

\bibitem{bi2015a}
Bi D, Henkes S, Daniels KE, Chakraborty B. 2015  The statistical physics of
  athermal materials. {\em Annu. Rev. Condens. Matter Phys.} \textbf{6},
  63--83.

\bibitem{degiuli2011}
DeGiuli E, McElwaine J. 2011  Laws of granular solids: Geometry and topology.
  {\em Phys. Rev. E} \textbf{84}, 041310.

\bibitem{degiuli2014}
DeGiuli E, Schoof C. 2014  On the granular stress-geometry equation. {\em EPL
  (Europhys. Lett.)} \textbf{105}, 28001.

\bibitem{satake1993}
Satake M. 1993  New formulation of graph-theoretical approach in the mechanics
  of granular materials. {\em Mech. Materials} \textbf{16}, 65--72.

\bibitem{fraternali2014}
Fraternali F, Carpentieri G. 2014  On the correspondence between 2{D} force
  networks and polyhedral stress functions. {\em Int. J. Space Struct.}
  \textbf{29}, 145--159.

\bibitem{desbrun2008}
Desbrun M, Kanso E, Tong Y. 2008  Discrete differential forms for computational
  modeling. In {\em Discrete differential geometry} pp. 287--324. Springer.

\bibitem{grady2010}
Grady LJ, Polimeni JR. 2010 {\em Discrete calculus: Applied analysis on graphs
  for computational science}.
Springer Science \& Business Media.

\bibitem{tonti2014}
Tonti E. 2014  Why starting from differential equations for computational
  physics?. {\em J. Comp. Phys.} \textbf{257}, 1260--1290.

\bibitem{lim2019}
Lim LH. 2019  Hodge {L}aplacians on graphs. {\em arXiv:1507.05379v4}.

\bibitem{ball2002}
Ball RC, Blumenfeld R. 2002  Stress field in granular systems: loop forces and
  potential formulation. {\em Phys. Rev. Lett.} \textbf{88}, 115505.

\bibitem{blumenfeld2003}
Blumenfeld R. 2003  Stress transmission in planar disordered solid foams. {\em
  J. Phys. A: Math. Gen.} \textbf{36}, 2399.

\bibitem{ishihara2012}
Ishihara S, Sugimura K. 2012  Bayesian inference of force dynamics during
  morphogenesis. {\em J. Theor. Biol.} \textbf{313}, 201--211.

\bibitem{redenbach2009}
Redenbach C. 2009  Microstructure models for cellular materials. {\em Comp.
  Mat. Sci.} \textbf{44}, 1397--1407.

\bibitem{desbrun2005}
Desbrun M, Hirani AN, Leok M, Marsden JE. 2005  Discrete exterior calculus.
  {\em arXiv preprint math/0508341}.

\bibitem{yavari2008}
Yavari A. 2008  On geometric discretization of elasticity. {\em Journal of
  Mathematical Physics} \textbf{49}, 022901.

\bibitem{ramola2017}
Ramola K, Chakraborty B. 2017  Stress response of granular systems. {\em J.
  Stat. Phys.} \textbf{169}, 1--17.

\bibitem{blanchard2019}
Blanchard GB, Fletcher AG, Schumacher LJ. 2019  The devil is in the mesoscale:
  mechanical and behavioural heterogeneity in collective cell movement. {\em
  Sem. Cell Dev. Biol.} \textbf{93}, 46--54.

\bibitem{blumenfeld2004}
Blumenfeld R. 2004  Stresses in isostatic granular systems and emergence of
  force chains. {\em Phys. Rev. Lett.} \textbf{93}, 108301.

\bibitem{noll2017}
Noll N, Mani M, Heemskerk I, Streichan SJ, Shraiman BI. 2017  Active tension
  network model suggests an exotic mechanical state realized in epithelial
  tissues. {\em Nature Physics} \textbf{13}, 1221.

\bibitem{noll2018}
Noll N, Streichan SJ, Shraiman BI. 2018  Geometry of epithelial cells provides
  a robust method for image based inference of stress within tissues. {\em
  arXiv preprint arXiv:1812.04678}.

\bibitem{spencer2017}
Spencer MA, Jabeen Z, Lubensky DK. 2017  Vertex stability and topological
  transitions in vertex models of foams and epithelia. {\em Eur. Phys. J. E}
  \textbf{40}.

\bibitem{barton2017}
Barton DL, Henkes S, Weijer CJ, Sknepnek R. 2017  Active vertex model for
  cell-resolution description of epithelial tissue mechanics. {\em PLoS Comp.
  Biol.} \textbf{13}, e1005569.

\end{thebibliography}

\end{document}